# 750 MHz radio frequency quadrupole with trapezoidal vanes for carbon ion therapy

Vittorio Bencini,[‡] Hermann W. Pommerenke[*,†] Alexej Grudiev, and Alessandra M. Lombardi
*European Organization for Nuclear Research (CERN), CH-1211 Geneva 23, Switzerland*

High-frequency linear accelerators are very suitable for carbon ion therapy, thanks to the reduced operational costs and the high beam quality with respect to synchrotrons, which are presently the only available technology for this application. In the framework of the development of a new linac for carbon ion therapy, this article describes the design of a compact 750 MHz radio frequency quadrupole (RFQ) with trapezoidal vanes. A new semianalytic approach to design the trapezoidal-vane RFQ is introduced together with the relevant beam dynamics properties. The RFQ is split into two decoupled rf cavities, both of which make use of a novel dipole detuning technique by means of length adjustment. The splitting is described both from the rf and the beam dynamics point of view. The paper concludes with the rf design of the full structure, including maximum surface field and thermal studies.

DOI: 10.1103/PhysRevAccelBeams.23.122003

## I. INTRODUCTION

The hadron therapy market is continuously expanding because of the advantages that this treatment brings with respect to conventional radiotherapy. One of the main limitations to the accessibility to hadron therapy is the cost of the accelerators needed to provide it. For this reason a lot of effort has been put into the development of compact and less expensive machines.

The idea of using high-frequency linear accelerators for hadron therapy was proposed in the early 1990s [1]. Together with the advances in the development of the accelerators, the interest in this research field has only grown since then. While proton therapy linacs have reached the industrialization phase—two machines are presently under construction or commissioning [2,3]—the same cannot be said for carbon ion machines, which are still at a conceptual design stage. Therefore, it is important to explore new possible solutions tailored to reduce size and costs of the accelerator and thought in a way that ease the transition between research and industrialization. Notable examples of the latest proposed designs are given by the Compact Carbon Ion Linac (ACCIL) [4] developed at Argonne National Laboratory, USA and the CABOTO machine (Carbon Booster for Therapy in Oncology) [5], whose latest version is described in Ref. [6].

One of the most recent proposals, fully described in Ref. [7], introduced the design of a 3 GHz "bent linac." The rationale of such a machine is to improve the aspect ratio of its footprint, studying a solution that could better fit into a hospital facility without affecting the treatment beam properties. In the low energy section of the machine, fully stripped carbon ions ($^{12}C^{6+}$) are produced by the ion source (TwinEBIS, operating at CERN [8]), transported in the low energy beam transport (LEBT) [9] and matched to the radio frequency quadrupole (RFQ), which represents the first rf accelerating structure. The RFQ has the critical role of shaping the beam in both the transverse and the longitudinal plane and prepare it for the injection in the following accelerating structures.

In order to reduce as much as possible the operational and building costs of the accelerator, all of its components have to be designed finding the best compromise between compactness, power consumption, and feasibility of construction. Driven by such requirements, a compact Carbon-RFQ operating at 750 MHz, a subharmonic of 3 GHz, has been developed at CERN. The proposed design (Fig. 1) follows the 750 MHz HF-RFQ for proton therapy, designed and built at CERN, that has been commissioned successfully in 2018 [10–14].

The rf design is based on the experience gained with the HF-RFQ and PIXE-RFQ [12,15–17]. However, with a length of 475 cm $\approx 12\lambda$ being more than twice as long as the HF-RFQ, the Carbon-RFQ had to be split into two separate rf cavities to ensure a stable field distribution. The length of each of the two fully decoupled cavities was used as a free parameter for dipole mode detuning, such that other dedicated detuning techniques are not necessary. This novel technique was conceptually proposed in Ref. [16] and has now been implemented for the first time. Two quadrupole vane extensions were used to rematch the beam across the intercavity drift.

[*]Corresponding author.
hermann.winrich.pommerenke@cern.ch
[†]Also at Institute of General Electrical Engineering, University of Rostock, D-18051 Rostock, Germany.
[‡]vittorio.bencini@cern.ch

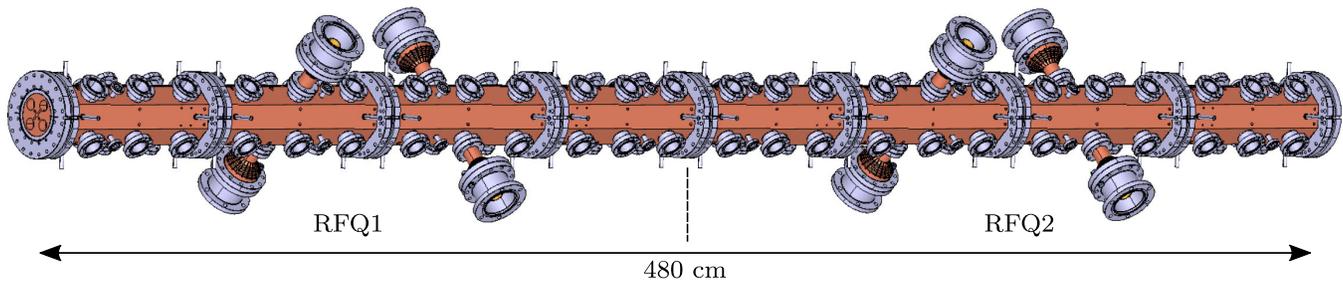

FIG. 1. Preliminary CAD model of the two Carbon-RFQ cavities RFQ1 and RFQ2. Each cavity consists of four individually brazed modules and features four input power couplers, 12 vacuum pumping ports, as well as 32 slug tuners.

The paper is structured as follows: Section II describes in detail the beam dynamics design, starting with the considered options for the standard-vane design (II A). It is followed by the transition to the trapezoidal design with the help of the sixteen-term potential function (II B), which is validated by means of particle tracking simulations (II C). In Sec. III, the splitting of the RFQ into two cavities is discussed from an rf perspective in terms of dipole detuning (III A) and rf decoupling (III B), as well as a beam dynamics point of view (III C). Finally, Sec. IV covers the remaining aspects of the rf design, including cavity shape (IV A), surface electric fields (IV B), auxiliaries and power couplers (IV C), as well as thermal simulation (IV D).

## II. BEAM DYNAMICS DESIGN

Section II discusses the two stages of the Carbon-RFQ beam dynamics design. First, the RFQ was developed with vanes whose longitudinal profile follows the two-term potential function. We denote these vane shape as standard vanes. Four main options were considered. Then, the final design choice was redesigned with trapezoidal vanes. An in-depth comparison by means of particle tracking follows.

### A. Design with standard vanes

#### 1. General design choices

The choice of the characteristic parameters used for the RFQ design was driven by various constraints. The $^{12}C^{6+}$-ion beam energy was defined by the maximum accelerating voltage that the LEBT can hold, in this case 30 kV, that leads to an input energy of 15 keV/u at the entrance of the RFQ. Concerning output energy, two different solutions at 2.5 MeV/u and 5 MeV/u were proposed in order to widen the range of options in the selection of the accelerating structure following the RFQ.

The frequency of the RFQ was chosen to be a subharmonic of the linac frequency (3 GHz) to allow the longitudinal beam injection. The specific frequency of 750 MHz was chosen as a compromise between the gain in structure length, which decreases as the frequency increases, the power consumption, which scales as $V_0^2 \cdot f^{3/2}$ [18], the beam acceptance, which decreases as the frequency increases, and the machinability. More detailed considerations on the frequency choice can be found in Ref. [11].

The minimum vane aperture $a$ (minimum distance between two opposite vanes) and the intervane voltage $V_0$ are strictly bound to the maximum surface field $E_{s,\max}$ that can be held by the vanes without breakdown. The Carbon-RFQ was designed aiming at $E_{s,\max} = 50.6$ MV/m, corresponding to 2.0 Kilpatrick [19,20]. Recently, the HF-RFQ for medical applications has been successfully commissioned [14], featuring the same frequency and maximum surface electric field, as well as a similar pulse length. As shown in Fig. 2, reducing the aperture $a$ results in an increasing surface field $E_{s,\max}$. At the same time, $E_{s,\max}$ can be reduced by decreasing the inter-vane voltage. Aperture and vane voltage are to be chosen finding the best compromise between power consumption ($P_0 \propto V_0^2$) and acceptance, which decreases with $a$, keeping $E_{s,\max}$ below the threshold. A summary of the Carbon-RFQ characteristic input parameters is shown in Table I.

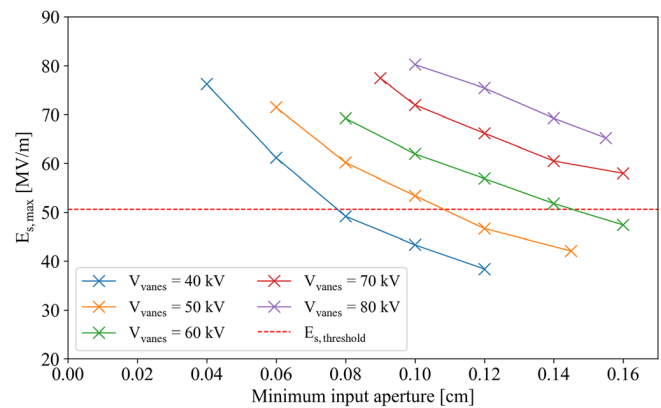

FIG. 2. Maximum surface field as a function of vane voltage and input aperture for different RFQ designs.

TABLE I. Carbon-RFQ general design parameters.

| Ion species | | $^{12}C^{6+}$ |
|---|---|---|
| Input energy | $\mathcal{W}_{in}$ | 15 keV/u |
| Output energy | $\mathcal{W}_{out}$ | 2.5 or **5.0** MeV/u |
| rf frequency | $f_0$ | 750 MHz |
| Inter-vane voltage | $V_0$ | 50 kV |
| Input current | | 0.19 mA |
| Input trn. emit. (90% norm.) | | 0.02 $\pi$mm mrad |
| Mid-cell aperture | $r_0$ | 1.411 mm |
| Minimum aperture | $a$ | 0.67 mm |
| Final synchronous phase | $\phi_s$ | −20 deg |
| Transverse curvature radius | $\varrho$ | 1.27 mm |
| Repetition rate | $f_{rep}$ | 200 Hz |
| Pulse length | $T_{pls}$ | 5 $\mu$s |
| Duty cycle | $d$ | 0.1% |

After the main RFQ parameters were fixed, the vane shape had to be defined. The choice of the vane profiles depends on a wide range of parameters that determine the beam dynamics in the RFQ. The main driver in the definition of modulation $m$, minimum aperture $a$, and synchronous phase $\phi_s$ along the structure is the kind of application the RFQ is tailored to. In the present work, two alternatives, a compact and a high-transmission design option, were assessed, each taking two possible output energies (2.5 MeV/u and 5 MeV/u) into account.

A comparison between the design options is shown in Table II, where the consequences of the different choices become clear in terms of particle transmission, length, rf power loss, and output longitudinal emittance. In the following, the main characteristics of each option are described in detail and the criteria used for the final choice are explained.

### 2. High-transmission RFQ

The vane parameters of the high-transmission RFQ were at first calculated using the LANL (Los Alamos National Laboratory) RFQ codes (Curli, RFQuick, Pari, and PARMTEQ) [21]. Once a baseline set of parameters was fixed, the vane profile was optimized further, increasing the maximum modulation and reshaping the vane profile in the gentle buncher (see Ref. [7] for more details). The function

$$m(z) = \frac{m_0 e^{rz_0} + m_1 e^{rz}}{e^{rz_0} + e^{rz}} \quad (1)$$

was used to describe the modulation profile in the gentle buncher, depending on the four parameters, $r$, $z_0$, $m_0$, and $m_1$. While $m_0$ and $m_1$ are fixed, corresponding to the modulation values at the entrance and exit of the gentle buncher, $r$ and $z_0$ can be optimized to minimize beam losses and $E_{s,max}$ along the gentle buncher. An example of the parametric function is shown in Fig. 3.

TABLE II. Values of transmission, length, power consumption and output longitudinal emittance for the considered RFQ design options with standard vanes. The final design choice is highlighted in bold face.

| | High transmission | | Compact | | |
|---|---|---|---|---|---|
| Output energy | 2.5 | **5** | 2.5 | **5** | MeV/u |
| Transmission | 99 | **99** | 43 | **43** | % |
| Length | 4.6 | **7.7** | 2.7 | **5.8** | m |
| Power | 470 | **790** | 280 | **600** | kW |
| $\varepsilon_{\phi\mathcal{W}}$ (rms normalized) | 0.4 | **0.4** | 0.14 | **0.14** | $\pi$deg MeV |

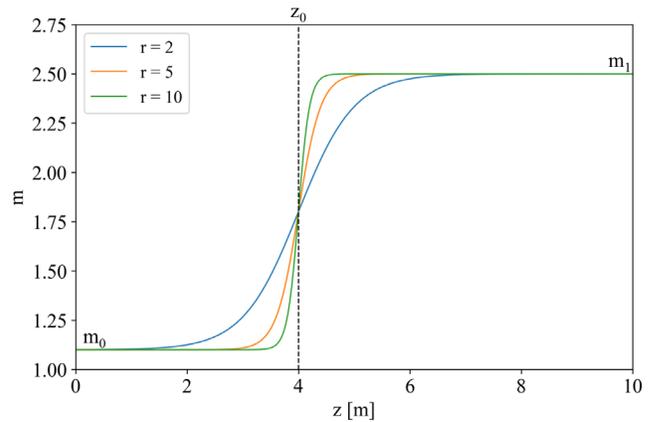

FIG. 3. Analytic function used to describe the modulation along the gentle buncher. $m_0$ and $m_1$ are the initial and final values of $m(z)$, $z_0$ shifts the function along z, $r$ defines the slope of the ramp.

The final $m$, $a$, and $\phi_s$ profiles resulting from the optimization are plotted in Fig. 4. The design algorithms used by the LANL codes are implemented to respect the adiabatic bunching condition and, moreover, to keep the bunch space-charge density constant along the RFQ, as detailed in Ref. [22]. In order to meet both conditions, the acceleration efficiency is kept very low in the first section of the RFQ (the shaper), that becomes the longest part of the 2.5 MeV/u variant, as can be seen in Fig. 4. The high transmission, therefore, comes at the price of a long structure and, consequently, high power consumption.

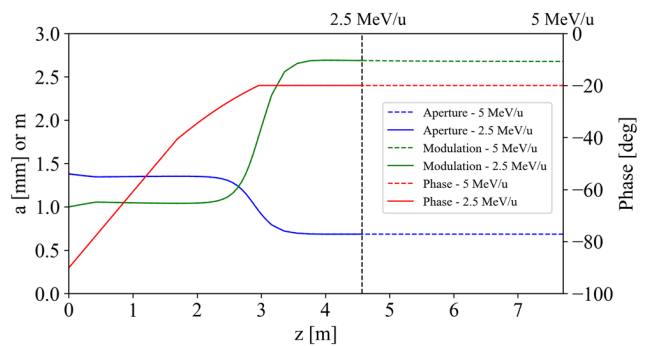

FIG. 4. Modulation, synchronous phase, and minimum aperture of the high transmission RFQ are plotted for the two final energies of 2.5 MeV/u (solid line) and 5 MeV/u (dashed line).

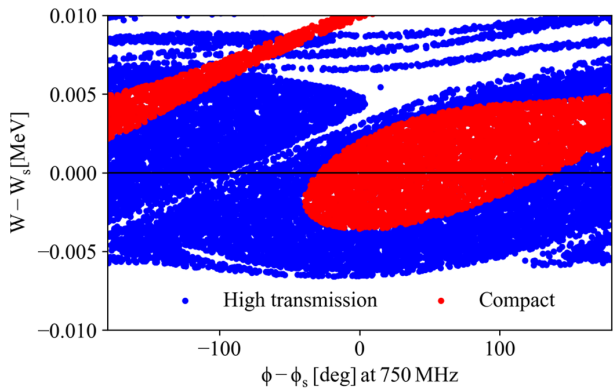

FIG. 5. Longitudinal acceptance for high transmission (blue) and compact (red) RFQ.

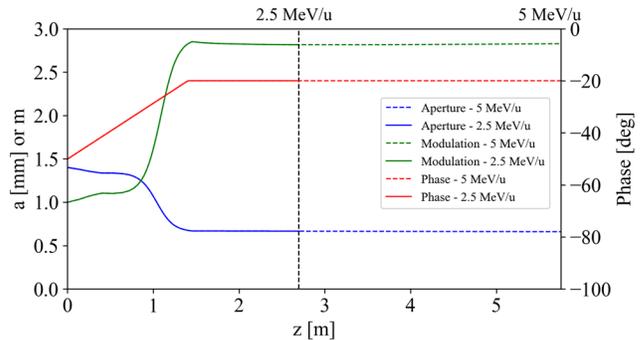

FIG. 6. Modulation, synchronous phase, and minimum aperture of the compact RFQ are plotted for the two final energies of 2.5 MeV/u (continuous line) and 5 MeV/u (dashed line).

### 3. Compact RFQ

The dimensions of the RFQ could be reduced by shortening the shaper and the gentle buncher, increasing the synchronous phase ramp-up. However, if the phase ramp was too steep, the capture process could not take place, resulting in uncontrolled losses. To overcome this issue, the synchronous phase at the entrance of the RFQ was increased from the conventional −90° to −50°. In this way, the phase increase is smooth enough to guarantee the beam bunching in the phase-energy phase space and the length of the shaper can be dramatically reduced. This technique, also implemented by HF-RFQ and PIXE-RFQ [11], provides a powerful tool to shape the beam on the longitudinal plane according to the requirements of the structure coming downstream the RFQ. The higher the input synchronous phase is, the lower is the output longitudinal emittance becomes. As a drawback, the particles that are not captured are lost, resulting in a decreased transmission.

As shown in Fig. 5, by increasing the phase, the longitudinal acceptance of the RFQ was decreased, meaning that about 50% of the particles entering the RFQ are not captured. Nevertheless, the amount of losses can was chosen by design and all the losses occur at injection energy, avoiding activation issues. The design of the gentle buncher of the compact RFQ was carried out using the same function described by Eq. (1). The modulation $m$, minimum aperture $a$, and synchronous phase $\phi_s$ profiles of the compact RFQ are plotted in Fig. 6.

### 4. Final design choice

To choose one of the proposed design options it is important to consider how carbon ion therapy treatment requirements translate to machine technical specifications. The treatment dose per unit volume and time usually amounts to 2 Gy/(minL), which corresponds to $4 \times 10^5$ carbon ions per pulse at $f_{rep} = 200$ Hz repetition rate. The amount of ions per pulse extracted by the TwinEBIS source is expected to be above $1 \times 10^9$ ions/pulse, much higher than a common treatment requirement. If a transmission of 50% is considered, the ion rate at the end of the RFQ would be $5 \times 10^8$ ions/pulse, thus still three order of magnitudes higher than the required one. The transmission of the RFQ is therefore not a critical concern in this specific application. Moreover, as shown in Table II, the phase-energy emittance at the end of the compact RFQ is more than two times smaller than for the high-transmission one, allowing for lossless bunch-to-bucket injection into a 3 GHz structure. For these reasons, the compact RFQ was chosen among the proposed options.

Two possible accelerating structures were considered to come downstream the RFQ. The first one consists of a 750 MHz IH-structure, described in Ref. [23]. This structure works at the same frequency as the RFQ, allowing for easy longitudinal injection, and can accept particles at 2.5 MeV/u, being reason why this energy was taken into account as a possible alternative. In the bent linac design, a 3 GHz SCDTL (side coupled drift tube linac) [24] was chosen after considerations discussed in Ref. [7]. The cell length in such structure equals $\ell_c = \beta\lambda$, where $\beta$ is the relativistic velocity and $\lambda$ the rf wavelength. At 3 GHz and 2.5 MeV/u, the cell length would be $\ell_c = 7.3$ mm, too short to be physically machined. At 5 MeV/u, $\ell_c = 10.3$ mm. An SCDTL structure with such cell length has already been machined and demonstrated to work on the LIGHT machine [25]. Thus, it was decided that the RFQ performs the acceleration up to 5 MeV/u.

### B. Design with trapezoidal vanes

The compact 5 MeV/u option arose as the final design choice from previous consideration. In the following, it was modified to feature trapezoidal vanes, increasing its acceleration efficiency.

#### 1. Trapezoidal vane geometry

A trapezoidal cell generally consists of two straight parts connected by a sinusoidal curve, providing a smooth longitudinal cell profile with a continuous first-order derivative. While other transitions may be used, we adopted the trapezoidal cell geometry proposed in Ref. [26]. From here on, the sinusoidal junction is denoted as the accelerating gap with length $g$. Furthermore, we let $r_0$ identify the mid-cell aperture, $a$ the minimum aperture, $m$ the modulation factor, and $L$ the cell length (Fig. 7). The longitudinal vane-tip profile follows

$$x(z) = \begin{cases} a & \text{for } 0 \leq z \leq \frac{L-g}{2} \\ r_0\left[1 + \frac{m-1}{m+1}\sin\frac{\pi}{g}\left(z - \frac{L}{2}\right)\right] & \text{for } \frac{L-g}{2} \leq z \leq \frac{L+g}{2}, \\ ma & \text{for } \frac{L+g}{2} \leq z \leq L \end{cases} \tag{2}$$

where $a = 2r_0/(m+1)$. The profile of the y-vane is mirrored with respect to the cell center at $z = L/2$. Note that for $g = L$, Eq. (2) just delivers the profile of a sinusoidal cell. The 3D geometry of the four-vane Carbon-RFQ trapezoidal cell with flat semicircular vanes is shown in Fig. 7. It is obtained by sweeping the transverse vane profile, a semi-circle with constant radius $\varrho = 0.9r_0$, along the trajectory given by the longitudinal profile.

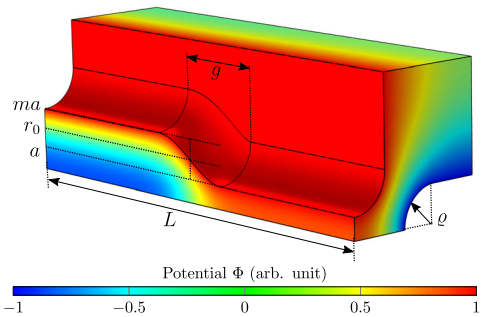

FIG. 7. Geometry and simulated electric potential of a trapezoidal RFQ cell.

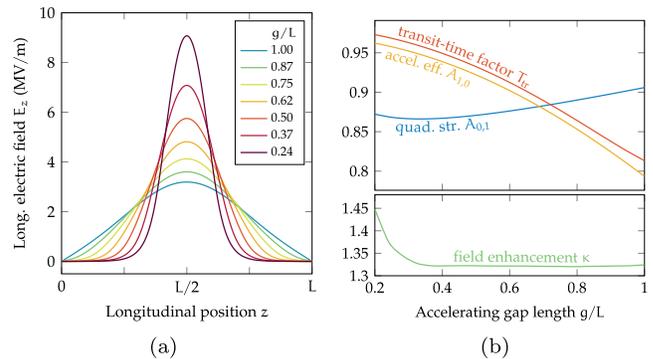

FIG. 8. Electric field $E_z$ along the beam axis ($r = 0$) of a trapezoidal cell with decreasing accelerating gap lengths $g/L$ (a), as well as quadrupole term $A_{0,1}$, acceleration efficiency $A_{1,0}$, and transit-time factor $T$ as functions of $g/L$ (b).

Figure 8(a) shows the longitudinal electric field $E_z$ along the beam axis of an exemplary trapezoidal cell ($m = 2.7$, $L = 10.5r_0$, $\varrho = 0.9r_0$) for varying normalized accelerating gap lengths $g/L$. While the spatial average of the peak accelerating field $E_0 = \langle E_z \rangle$ stays virtually constant, the transit-time factor $T$ of the synchronous particle [27] is increased. For an ideal standard cell following the two-term potential, $T = \pi/4 \approx 0.786$. While for purely sinusoidal cells ($g/L = 1$) the transit-time factor is already slightly higher, it could be increased to $T = 0.988$ by decreasing $g/L$ within the cell parameter space of the Carbon-RFQ. This value is close to the theoretical limit of $T = 1$. The increase in $T$ is equivalent to an increase in acceleration efficiency $A_{1,0}$. Figure 8(b) shows the quantities as functions of $g/L$. Generally, the increased acceleration comes at the price of a smaller quadrupole focusing term $A_{0,1}$. This is also the case for sinusoidal cells when increasing the modulation $m$—but not for RFQs featuring standard two-term potential cells, where the focusing is generally kept constant.

Another important figure is the maximum surface electric field $E_{s,\text{max}}$, representing one of the major performance limits for any linac [27]. The lower plot of Fig. 8(b) depicts this quantity as a function of $g/L$, expressed as the field enhancement $\kappa = E_s/(V_0/r_0)$. Figure 9 visualizes the maximum field region as the accelerating gap is decreased: the maximum field increases, and its location changes from a widespread area in concave region of the electrode surface (a) to a more localized hot spot in the convex region (d). Although the increased field poses a performance limit to accelerating cells located toward the end of an RFQ, such as the one studied here, the value is significantly surpassed by the short cells found in the gentle buncher during the modulation ramp-up (see also

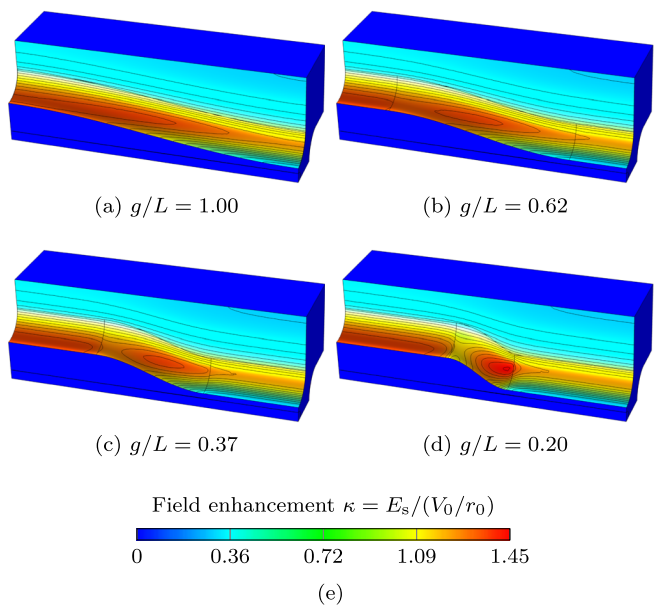

FIG. 9. Surface electric field in a trapezoidal cell, shown for varying gap lengths $g/L$ (a)–(d), and (e) legend.

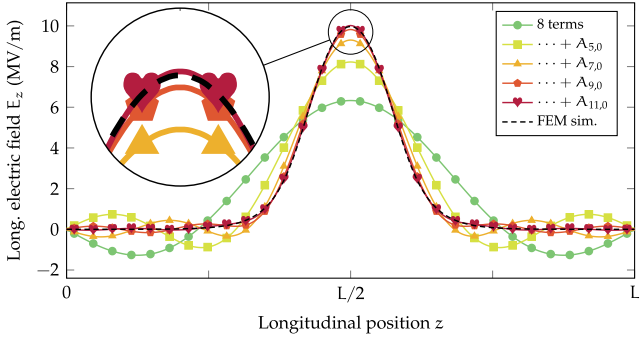

FIG. 10. Approximation of the electric field simulated in COMSOL (black dashed line) on the beam axis of a trapezoidal cell by the potential function with increasing number of terms.

Figs. 11(d) and 13). The bottleneck thus remains in the upstream stages of the RFQ.

### 2. Sixteen-term potential function

The electric field of trapezoidal vanes cannot be described by the two-term or eight-term potential function used by the available RFQ design codes [21,27–31], PARMTEQ [21] being the most common. Previously, 3D simulation code was employed to design RFQs with trapezoidal electrodes, taking the electrode geometry directly into account [26,32–37]. While this procedure delivers highly accurate results, it is also time-consuming, as for each parameter change new simulations have to be performed to solve the Laplace equation.

In this paper, we augment the well-known eight-term potential function by higher-order terms, accounting for the particular field pattern of the trapezoidal electrode. A similar approach has recently been published in Ref. [38]. Figure 10 shows the approximation of the simulated on-axis electric field $E_z$ by the potential function. By adding the longitudinal coefficients $A_{5,0}$, $A_{7,0}$, $A_{9,0}$, and $A_{11,0}$, the approximation error falls below 1% of the peak field. Additionally, the four coefficients $A_{4,1}$, $A_{6,1}$, $A_{4,3}$, and $A_{6,3}$ are included in order to accurately describe possible distortions of the transverse quadrupole field. The full sixteen-term potential reads

$$\frac{\Phi(r,\vartheta,z)}{V_0/2} = A_{0,1}\left(\frac{r}{r_0}\right)^2 \cos(2\vartheta) + A_{0,3}\left(\frac{r}{r_0}\right)^6 \cos(6\vartheta) + A_{1,0}\mathcal{I}_0(kr)\cos(kz) + A_{3,0}\mathcal{I}_0(3kr)\cos(3kz)$$
$$+ A_{5,0}\mathcal{I}_0(5kr)\cos(5kz) + A_{7,0}\mathcal{I}_0(7kr)\cos(7kz) + A_{9,0}\mathcal{I}_0(9kr)\cos(9kz) + A_{11,0}\mathcal{I}_0(11kr)\cos(11kz)$$
$$+ A_{1,2}\varphi_{1,2}(\mathbf{r}) + A_{3,2}\varphi_{3,2}(\mathbf{r}) + A_{2,1}\varphi_{2,1}(\mathbf{r}) + A_{2,3}\varphi_{2,3}(\mathbf{r}) + A_{4,1}\varphi_{4,1}(\mathbf{r}) + A_{4,3}\varphi_{4,3}(\mathbf{r}) + A_{6,1}\varphi_{6,1}(\mathbf{r}) + A_{6,3}\varphi_{6,3}(\mathbf{r}),$$
(3)

with $\varphi_{\mu,\nu}(\mathbf{r}) = \mathcal{I}_{2\nu}(\mu kr)\cos(2\nu\vartheta)\cos(\mu kz)$, where $\mathcal{I}_n(\cdot)$ is the modified Bessel function of first kind and $n$th order.

Multipole coefficients and maximum surface field are precomputed and stored in a lookup table similar to PARMTEQ [21,28] for a discrete number of cells in the parameter space of interest. The coefficients for a specific cell geometry are then obtained by interpolation. In general, the trapezoidal cell of a four-vane RFQ with transverse semicircular vane tips (Fig. 7) is described by six independent parameters: $L$, $m$, $g$, $\varrho$, $r_0$, and $V_0$, all of which may vary along the RFQ. For the computation of the potential field and its coefficients, however, $V_0$ merely appears as a scalar factor, while length quantities can be replaced by the dimensionless parameters $L/r_0$, $g/L$, and $\varrho/r_0$. Furthermore, for the Carbon-RFQ, the transverse curvature radius $\varrho$ is constant. The remaining three free parameters were swept over the ranges $0.75 \leq L/r_0 \leq 20$, $0.15 \leq g/L \leq 1$, and $1 \leq m \leq 3.2$. A limit to the parameter space is given by the longitudinal radius of curvature $\varrho_L$:

$$\varrho_{L,\min} \leq \varrho_L = \frac{g^2}{\pi^2 r_0}\left(\frac{m+1}{m-1}\right) \quad (4)$$

The theoretical lower limit is given by the transverse radius of curvature, $\varrho_{L,\min} = \varrho = 0.9r_0$, which must be smaller than the longitudinal radius of curvature in order to construct a physically possible geometry. In practice however, $\varrho_{L,\min}$ is limited by the radius of the cutter tool used to machine the RFQ vanes. Thus, the cuboid spanned by the parameter ranges is incomplete starting from the corner corresponding to short cells with short accelerating gaps and large modulations. In total, 2280 different cell geometries were precomputed for the Carbon-RFQ using the finite element method (FEM) Laplace equation solver of COMSOL Multiphysics® [39].

Figure 11 shows a selection of the computed multipoles over the parameter space. Most important for the beam dynamics design are the quadrupole term $A_{0,1}$ (a) and the acceleration efficiency $A_{1,0}$ (b). The plots highlight that decreasing $g/L$ has in first approximation the same effect as increasing $m$—increasing acceleration at the expense of focusing. This equivalence can also be seen from the newly introduced coefficient $A_{5,0}$ (c), which also highlights the importance of the newly introduced higher-order terms for cells with $g/L \ll 1$.

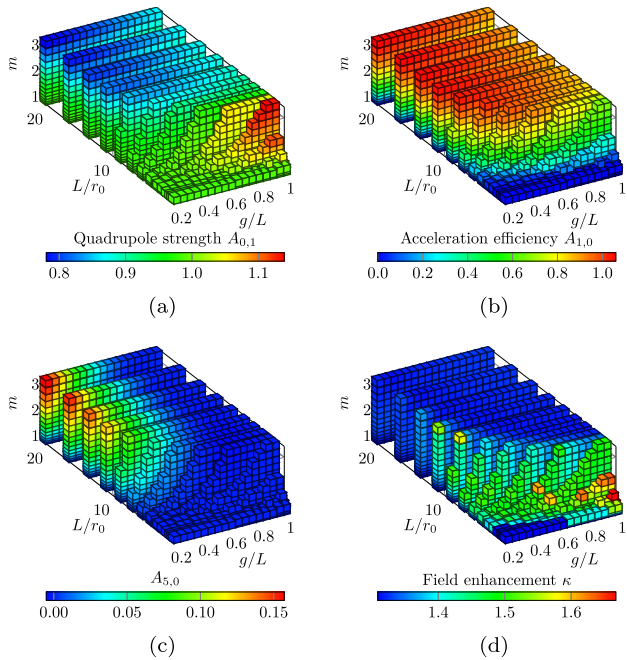

FIG. 11. Computed multipole coefficients as functions of cell length $L/r_0$, modulation $m$, and gap length $g/L$ over the parameter space available to the Carbon-RFQ cells, where $\varrho/r_0 = 0.9$. Shown are the quadrupole term $A_{0,1}$ (a), the acceleration efficiency $A_{1,0}$ (b), a new higher-order longitudinal term $A_{5,0}$ (c), and the field enhancement $\kappa$ (d).

Finally, Fig. 11(d) shows the field enhancement as a function of the three parameters. Here it becomes clear that—for a constant $V_0$ and $r_0$—one should expect the maximum surface field in the short cells in the bunching section at the beginning of the RFQ. A decrease in $g/L$ raises the field only slightly. This was confirmed during the rf design (Section IV B, Fig. 27). One can take advantage of this fact by designing an RFQ with nonuniform $V_0$ and $r_0$, for example by increasing the field amplitude toward the high-energy end of the RFQ, evenly distributing the maximum surface field strength. While this technique has occasionally been used in RFQs featuring standard or sinusoidal electrodes, it seems particularly useful for trapezoidal-electrode RFQs with their highly efficient accelerator sections that still feature comparatively low field enhancement. However, the Carbon-RFQ was realized with constant voltage, mid-cell aperture, and transverse radius for simplified rf design and construction.

### 3. Cell-by-cell design

The Carbon-RFQ with trapezoidal vanes was designed using a simple cell-by-cell algorithm respecting the linac synchronism conditions [27]: The cell length is given by

$$L = \frac{\langle \beta \rangle \lambda}{2}, \qquad \langle \beta \rangle = \frac{\beta_{\text{in}} + \beta_{\text{out}}}{2}, \tag{5}$$

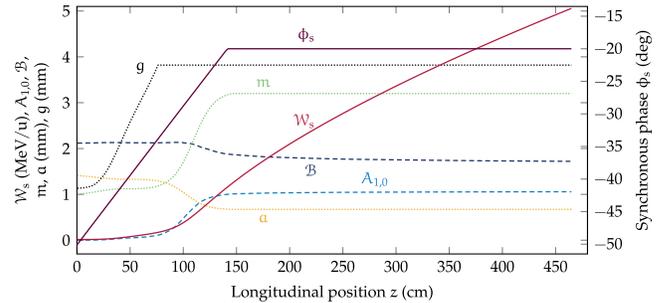

FIG. 12. Beam dynamics parameters of the Carbon-RFQ with trapezoidal vanes.

where $\beta_{\text{in}}$ and $\beta_{\text{out}}$ denote the velocity of the synchronous particle at cell entrance and cell exit, respectively. The synchronous energy gain upon traversing the cell amounts to

$$\mathcal{W}_{\text{out}} = \mathcal{W}_{\text{in}} + \frac{\pi}{4} q V_0 A_{1,0}\left(\frac{L}{r_0}, \frac{g}{L}, m\right) \mathcal{I}_0(kr) \cos\phi_s. \tag{6}$$

The trapezoidal design is a modification of the compact 5 MeV/u RFQ devised using the PARMTEQ codes with standard-vane geometry based on the two-term potential function (Sec. II A 3, Fig. 6). From this design, the same synchronous phase $\phi_s(z)$ and modulation $m(z)$ provided the algorithm input.[1] With $r_0$ and $V_0$ constant, and $k = \pi/L$, the remaining free parameter is the accelerating gap length $g$. Optimizing the RFQ for maximum acceleration efficiency, $g$ was chosen as small as possible while maintaining a minimum longitudinal curvature radius of $\varrho_{L,\min} = 2$ mm. With Eq. (4), $g$ is given by

$$g = \min\left(L, \pi\sqrt{\varrho_{L,\min} r_0 \frac{m-1}{m+1}}\right), \tag{7}$$

as $g/L$ must not exceed unity. Equations (5), (6), and (7) form a nonlinear system that is easily solved iteratively, converging within a few iterations. The determined output energy $\mathcal{W}_{\text{out}}$ is then used as input energy $\mathcal{W}_{\text{in}}$ for the subsequent cell, and the procedure is repeated until the required output energy is reached.

Emphasis is placed on the fact that the acceleration efficiency $A_{1,0}$ in Eq. (6) is obtained by multidimensional linear interpolation from the precomputed lookup table [Fig. 11(b)] for a given geometry, avoiding repetitive electrostatic simulation. Using this technique, the vane design required merely a few seconds. Furthermore, an electroquasistatic field map can be produced rapidly based on the sixteen-term potential function, which allows for particle tracking in a tool of choice to validate the design.

Figure 12 shows the beam dynamics parameters of the finalized RFQ with trapezoidal vanes. Instead of

---

[1]The modulation $m$ was adjusted such that with the different cell geometry the same minimum aperture $a$ and mid-cell aperture $r_0$ are achieved. This is illustrated in Sec. II B 4.

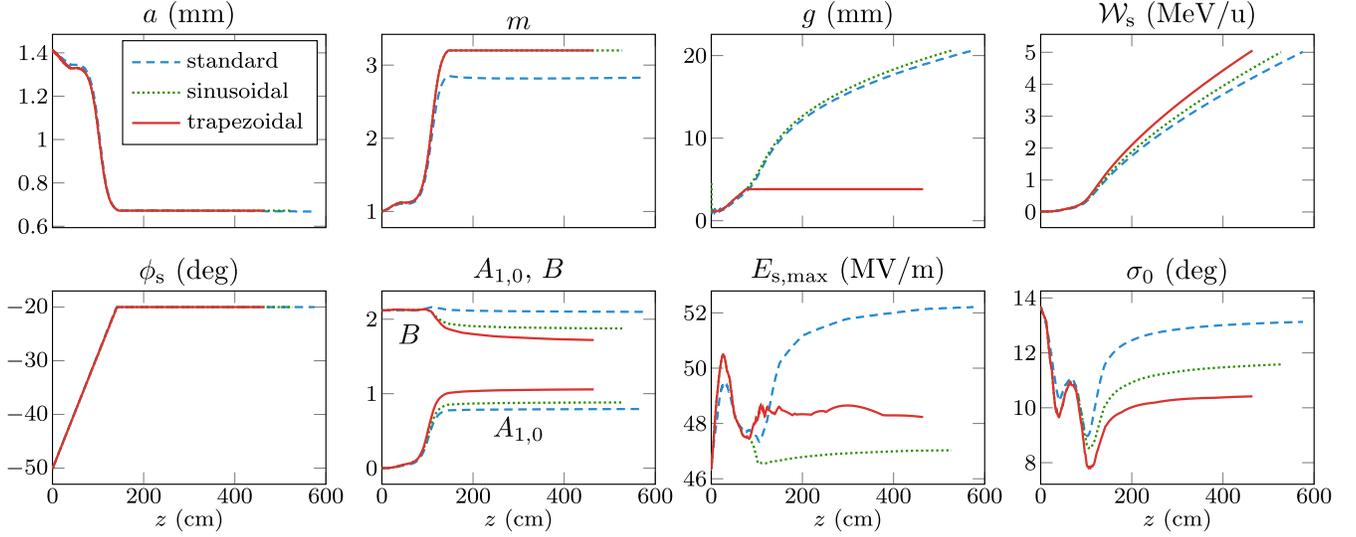

FIG. 13. Comparison between standard-vane RFQ, trapezoidal-vane RFQ, and a design employing purely sinusoidal vanes, for the same final energy of 5 MeV/u. Parameters are shown as functions of the position $z$, while $V_0$, $r_0$, and $\varrho$ are constant and equal in all designs. The gap length $g$ is a free parameter for the trapezoidal design, but equals the cell length in the other designs.

the quadrupole term $A_{0,1}$, we depict the effective focusing force

$$B = \frac{q}{m_0 c^2} \lambda^2 \frac{A_{0,1} V_0}{r_0^2}. \quad (8)$$

### 4. Design analysis

A comparison between the most relevant parameters of the standard-vane and trapezoidal-vane RFQ is shown in Fig. 13. Additionally, the parameters of a purely sinusoidal design are shown, corresponding to the special case of the trapezoidal geometry where $g/L = 1$. This allows to assess which differences solely originate in changing from two-term potential cells to sinusoidal cells, and which differences come from introducing the straight vane parts to the sinusoidal cell, i.e., choosing $g/L < 1$.

First, it is important to note that the modulation $m$ of trapezoidal and sinusoidal vanes is higher than that of the standard vane. The reason comes from the definition of the geometrical parameters of the RFQ cell. For the first two cases, the mid-cell aperture amounts to $r_0 = a(m+1)/2$, while for standard vanes, which originate directly from the two-term potential function, it is given as $r_0 = aX^{-1/2}$ [27], where

$$X = \frac{\mathcal{I}_0(ka) + \mathcal{I}_0(mka)}{m^2 \mathcal{I}_0(ka) + \mathcal{I}_0(mka)}. \quad (9)$$

Therefore, in order to keep $r_0$ and the minimum aperture $a$ equal to the original design, $m$ had to be increased, leading to the difference in Fig. 13.

The larger modulation directs more electric field in longitudinal direction, increasing acceleration efficiency $A_{1,0}$ (Fig. 13). Consequently, choosing sinusoidal over standard vanes already leads to a shorter RFQ. This is only amplified by reducing the accelerating gap length, such that the trapezoidal vanes allow for reaching the final energy with a shorter—and thus less power consuming—cavity. The increased acceleration efficiency comes at the expense of focusing strength $B$ in the accelerator section of the RFQ where the modulation is maximum. The consequence is a non-negligible reduction in transverse phase advance $\sigma_0$ [27]. The trapezoidal geometry also changes the distribution of the maximum surface electric field $E_{s,\max}$ on the vanes, which results to be lower than for standard vanes despite the smaller longitudinal curvature radius.

### C. Particle tracking results

The beam dynamics design was validated by means of particle tracking simulation. The input distribution, as expected from the LEBT, was tracked into field maps of both the standard-vane and the trapezoidal-vane RFQ. The tracking was performed using two different codes (TRAVEL [40] and RF-TRACK [41]) to cross-check the results.

The RFQ was designed to accept $1 \times 10^9$ ions in a pulse of 5 μs, which translates to a nominal pulse current of 0.19 mA. In Ref. [7], particle tracking was performed for beams with different currents, and it was highlighted that the space charge effects for the nominal current are negligible.

Convergence studies have been carried out with respect to spatial resolution of the field maps and time resolution of the tracking. The phase space parameters converged with a relative error of 10% or better for a field map step size of 50 μm (15 transverse samples between beam axis and minimum aperture) and a time resolution of 30 steps per RFQ cell.

TABLE III. Carbon-RFQ output phase space parameters for conventional and trapezoidal vanes.

| | | Standard vanes | | Trapezoidal vanes | | |
| --- | --- | --- | --- | --- | --- | --- |
| | | 8-term | COMSOL FEM | 16-term | COMSOL FEM | |
| Transmission | $\mathcal{T}$ | 45.96 | 44.99 | 48.82 | 48.31 | % |
| Energy | $\mathcal{W}_{\text{out}}$ | 5.021 | 5.018 | 5.015 | 5.014 | MeV/u |
| Synchronous phase | $\phi_s$ | −19.7 | −20.1 | −19.5 | −19.1 | deg |
| Horizontal normalized rms emittance | $\varepsilon_{xx'}$ | 0.0218 | 0.0197 | 0.0183 | 0.0184 | $\pi$ mm mrad |
| Vertical normalized rms emittance | $\varepsilon_{yy'}$ | 0.0214 | 0.0202 | 0.0178 | 0.0181 | $\pi$ mm mrad |
| Longitudinal normalized rms emittance | $\varepsilon_{\phi\mathcal{W}}$ | 0.1353 | 0.1264 | 0.1045 | 0.1448 | $\pi$ deg MeV |

### 1. Semianalytic vs. FEM-based field map

Both for the standard-vane and the trapezoidal-vane RFQ, two field maps were used: the first was built using the eight-term or sixteen-term potential function defined in Sec. II B 2, where the physical vane geometry is not included and the aperture is approximated by a circle with radius $r_0$ along the entire RFQ. The second map is the result of an FEM electrostatic simulation performed with COMSOL Multiphysics® [39], and takes the physical vane geometry into account.

In Table III and Fig. 14, the beam distributions resulting from the tracking of 10 000 macroparticles into the four maps are compared. Noticeable for both designs is that the transmission was slightly lower when the COMSOL FEM map was used instead of the potential map. This can be explained by the different aperture definition in the two cases: From Fig. 14, it can be seen that for the FEM map tracking the halo around the transverse phase space is less spread and defined by a more net line. The reason is found in the transverse acceptance defined by the mid-cell aperture $r_0$, being slightly bigger than the one defined by the physical vane geometry, thus allowing transmission of the more exterior particles of the distribution that are cut by the physical vane. Despite these observations, it is noted that the differences in the transverse plane are in the order of 1% and the corresponding rms emittances show very close agreement.

In the longitudinal plane, larger discrepancies can be found. Although the potential function is a very good approximation of the axial accelerating field, differences of few percent remain with respect to the COMSOL FEM map, as indicated in Sec. II B 2. Additional, much smaller deviations may originate in the cell-to-cell tapering of the cell parameters, that has to be implemented to obtain a smooth electrode surface, but is only approximately considered by the potential function field map. Even though the average final energy and synchronous phase are nearly identical, the output particle distributions differ slightly in shape and thus in rms emittance. This effect is more marked for the trapezoidal vanes, where the difference between potential and FEM map is larger than in the case of the standard vane design. However, the discrepancies are small enough: all four simulated longitudinal phase spaces (Fig. 14) fulfill well the requirements given by the SCDTL acceptance [7]. Thus, the sixteen-term potential function represents a highly useful tool for rapidly designing the trapezoidal-vane RFQ. Nevertheless, the design should always be confirmed by means of FEM simulation of the real geometry.

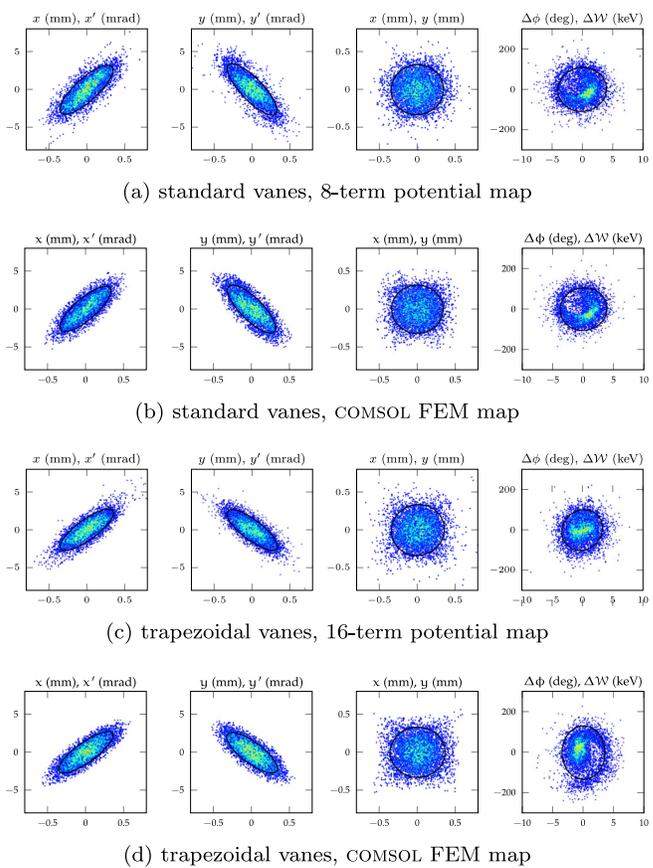

FIG. 14. Comparison of the output phase spaces of the standard (a), (b) and trapezoidal vane design (c), (d), each shown for the potential function and FEM field maps.

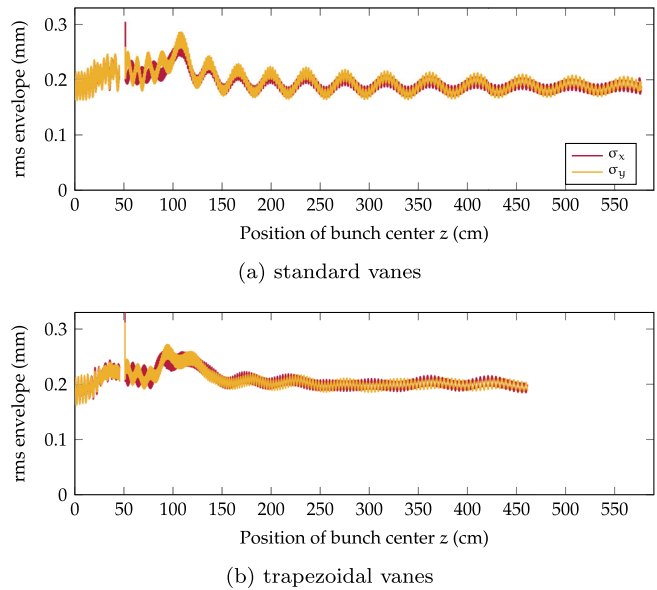

FIG. 15. Rms beam envelopes computed from the FEM field map for the standard (a) and trapezoidal design (b). (The discontinuity at $z = 50$ cm is a nonphysical artifact originating in the removal of macroparticles from the tracking simulation, that are not captured in the rf bucket.)

#### 2. Standard-vane vs. trapezoidal-vane RFQ

Figure 15 shows the rms beam envelopes of standard-vane and trapezoidal-vane RFQ. The corresponding output phase spaces are depicted in Figs. 14(b) and 14(d). As expected, the envelopes are nearly identical in the first 90 cm of the RFQ, and then start to differ. This occurs at the same point in $z$ where the profiles of effective focusing strength $B$, acceleration efficiency $A_{1,0}$, and consequently of the transverse phase advance $\sigma_0$, for the standard-vane and trapezoidal-vane RFQs start to differ (Fig. 13).

As mentioned in Sec. II A 3, one of the main advantages of the beam dynamics design choices adopted for the compact RFQ—with both standard or trapezoidal vanes—is that all the losses occur at injection energy, preventing activation issues. The particle tracking in the COMSOL FEM maps allows for recording the actual losses on the vanes. The results are summarized in Fig. 16, where the percentage of lost particles as a function of the particle energy is plotted for both designs. It is noticed that almost the totality of the losses is concentrated in the first bin, which corresponds to the injection energy.

### III. SPLITTING INTO TWO CAVITIES

With a length of 475 cm $\approx 12\lambda$, the Carbon-RFQ had to be split into two separate rf cavities to ensure a stable field distribution. Several (eight) independent rf power sources are planned to be used to feed the RFQ. Thus, instead of commonly used coupling cells, the cavities are fully decoupled from the rf perspective, but share vacuum. By this, both cavities may be designed, tuned, commissioned, and maintained separately. Moreover, field amplitude and phase may be adjusted independently.

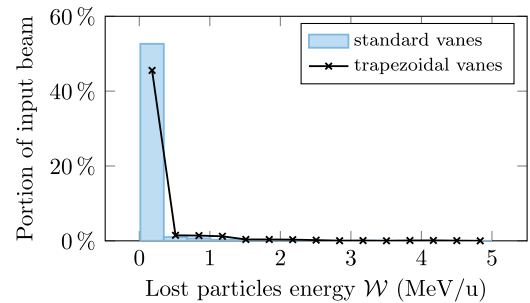

FIG. 16. Energy distribution of the particles lost during the particle tracking through the FEM field map of both designs.

Designing the transition required close cooperation between beam dynamics and rf design, namely, choosing the appropriate cavity lengths for dipole detuning (Sec. III A), and rematching the beam across the intercavity drift (III C). Thus, we discuss it here in a separate section, while remaining aspects of rf design are covered in Sec. IV.

#### A. Dipole-mode detuning by length adjustment

In any four-vane RFQ, the frequency separation between the operating $TE_{210}$ mode and the closest dipole modes should be maximized to increase field stability and robustness against machining errors [27], which is commonly referred to as dipole-mode detuning. For both RFQ1 and RFQ2, the detuning was achieved solely by adjusting the cavity length, a novel technique proposed in Ref. [16]. Avoiding the use of any dedicated geometric features, rf power losses were reduced and the machining was simplified.

#### 1. Transmission line model

A transmission line model (TLM) was used to calculate quadrupole and dipole frequencies in order to rapidly estimate the required length. The TLM is shown in Fig. 17, and the equivalent matrix equation reads

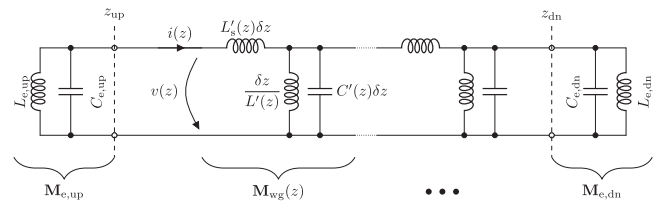

FIG. 17. Transmission line model of the symmetric four-vane RFQ cavity with ends, valid for both quadrupole and dipole modes.

$$\begin{bmatrix} v(L_{\mathrm{RFQ}}) \\ v'(L_{\mathrm{RFQ}}) \end{bmatrix} = \underbrace{\mathbf{M}_{e,\mathrm{dn}} \left( \prod_{z=z_{\mathrm{up}}}^{z_{\mathrm{dn}}} \mathbf{M}_{\mathrm{wg}}(z) \right) \mathbf{M}_{e,\mathrm{up}}}_{\mathbf{M}_{\mathrm{RFQ}}} \begin{bmatrix} v(0) \\ v'(0) \end{bmatrix}. \quad (10)$$

Here, $v(z)$ is the voltage profile along the RFQ, and

$$\mathbf{M}_{\mathrm{wg}}(z) = \begin{bmatrix} \cos k(z)\delta z & 1/k(z) \sin k(z)\delta z \\ -k(z) \sin k(z)\delta z & \cos k(z)\delta z \end{bmatrix} \quad (11)$$

models the mode propagation along the RFQ obeying the dispersive wave equation [27,42]. The position-dependent wave number $k^2(z) = (\omega/c)^2 - k_{\mathrm{co}}^2(z)$, where $k_{\mathrm{co}}(z) = 1/[c^2 L'(z) C'(z)]$, is obtained from the lumped circuit parameters by simulating thin RFQ slices (Fig. 25, second and third plot). The RFQ is considered as an inhomogeneous waveguide consisting of many homogeneous slices by multiplying the corresponding matrices. The 3D end geometries are modeled by additional matrices [42]

$$\mathbf{M}_e = \begin{bmatrix} 1 & 0 \\ \dfrac{1}{c^2 C'} \left( \dfrac{1}{L_e} - \omega^2 C_e \right) & 1 \end{bmatrix} \quad (12)$$

describing shunt LC circuits. $C_e$ and $L_e$ depend on the vane nose and on the shape of the vane undercut window, respectively. The parameters are chosen such that the ends appear as perfect open-circuit boundaries to the $TE_{210}$ operating mode. On all other modes they act as mismatched impedances that perturb frequency and field distribution. The specific values of $C_e$ and $L_e$ are determined by 3D eigenmode simulations of short end segments. The eigenmode frequencies are found by (numerically) solving $[\mathbf{M}_{\mathrm{RFQ}}]_{2,1} = 0$, i.e. finding the frequencies for which the longitudinal current $v'(L_{\mathrm{RFQ}})$ at the end of the RFQ vanishes for any given voltage $v(0)$.

#### 2. Resulting spectra

The TLM was used to estimate the required length of the two RFQ cavities in order to maximize the spectral margin between the operating $TE_{210}$ mode and the closest dipole modes. The length of RFQ1 was determined by the plane of splitting, which can be freely positioned at the end of any accelerating cell. The remaining accelerating cells give a preliminary length for RFQ2. Figure 18 shows how the frequencies of the eigenmodes of RFQ1 change with the position of the splitting plane. For the $TE_{114}$ mode—the higher of the two most critical dipole modes—the sensitivity reads approximately 0.6 MHz/cell. Thus, the placement of dipole modes can be precisely controlled by varying the overall RFQ length in steps of the cell length, or $\beta\lambda/2$, where $\beta$ is the synchronous particle velocity at the plane of splitting.

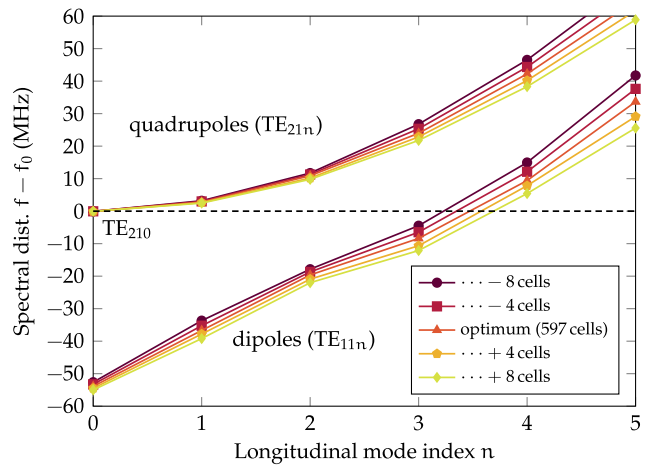

FIG. 18. Sensitivity of the eigenmode spectrum of RFQ1 with respect to its length, given as the number of accelerating cells.

We chose to split the RFQ such that similar lengths, and thus similar rf and mechanical designs arose for both cavities. This positions the $TE_{210}$ mode between $TE_{113}$ and $TE_{114}$ (Fig. 19). As the preliminary length of RFQ2 did not result in an acceptable spectrum, additional accelerating

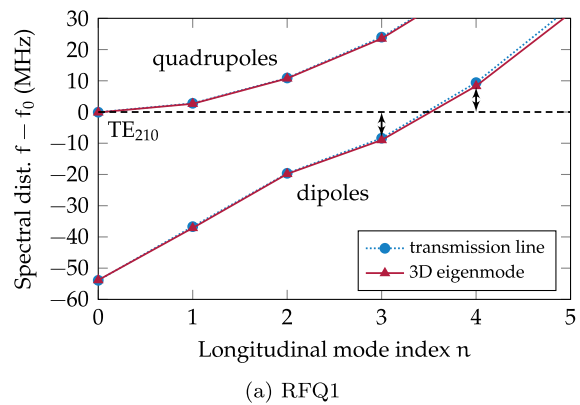

(a) RFQ1

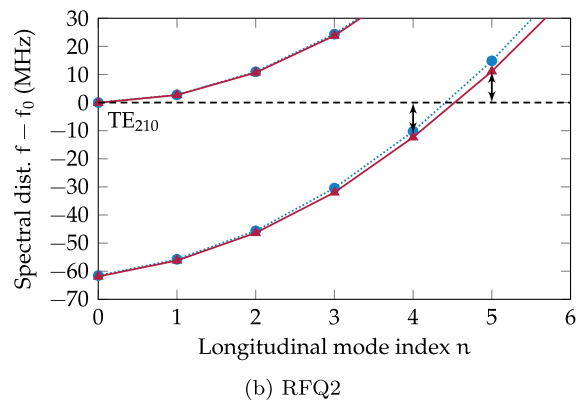

(b) RFQ2

FIG. 19. Eigenmode spectra of the two cavities RFQ1 (a) and RFQ2 (b) after dipole detuning using the RFQ length. The margins between $TE_{210}$ and the closest dipole modes read ±8 MHz and ±11 MHz, respectively.

cells were appended to shift the $TE_{210}$ mode symmetrically between $TE_{114}$ and $TE_{115}$.

Passing on to eigenmode simulations performed in CST Microwave Studio® [43] for an initial 3D model, we noticed that the dipole mode frequencies were slightly lower than those calculated from the TLM. The deviations of the dipole modes closest to the $TE_{210}$ mode between TLM and eigenmode simulation read approximately 1 MHz for RFQ1 and 2 MHz to 3 MHz for RFQ2. While the relative error in frequency is less than 0.5%, it corresponds to an error in length of roughly two RFQ cells. Thus, the 3D models were adjusted in a second iteration. Figure 19 shows the spectra of the finalized RFQ1 (a) and RFQ2 (b) cavities. Both 3D simulation and TLM results of the second iteration are depicted. The spectral margins between $TE_{210}$ mode and the closest dipole modes approximately read ±8 MHz for RFQ1, which features a length of 235 cm and 597 accelerating cells. In RFQ2, ±11 MHz mode separation was obtained by appending two accelerating cells, increasing its output energy from 5.02 MeV/u to 5.06 MeV/u and its length to 233 cm (129 cells). The quadrupole mode margin (distance between $TE_{210}$ and $TE_{211}$) amounts to 2.7 MHz in both cavities.

### B. Decoupling of RFQ cavities

In order to tune and even commission RFQ1 and RFQ2 separately, both cavities require their own end plates including bead-pull holes with 24 mm diameter. For each of the face-to-face mounted end plates between the cavities, a minimum thickness of approximately 2 cm is required. This value arises from mechanical considerations regarding manufacturing, stability, and cooling [44]. The combined thickness of ≈ 4 cm is sufficient to decouple the operating modes of both cavities, i.e., that the field coupling through the bead-pull holes is negligibly small.

Consider only one eigenmode, the $TE_{210}$ operating mode with frequency $f_0$ in each RFQ cavity. In a structure composed of two coupled resonators, two orthogonal modes can be excited: the zero-mode with a lower frequency $f_{PMC} < f_0$ and the π-mode with a higher frequency $f_{PEC} > f_0$. The two cases correspond to perfect magnetic (PMC) or perfect electric (PEC) boundary conditions at the coupling plane, respectively. The coupling is proportional to the bandwidth $f_{PEC} - f_{PMC}$ normalized to $f_0$. For zero coupling, both frequencies equal $f_0$ and the bandwidth vanishes. The bandwidth can be obtained from an eigenmode simulation of a short end segment of length $\Delta L$, where either the PEC or PMC boundary condition is applied to the face of the bead-pull hole, delivering the respective frequencies. Figure 20(a) visualizes the magnetic field $H$ in the end segment, shown here with an end plate thickness of $d_e = 2$ cm for the PMC case.

In order to obtain a quantity suitable for comparison, an effective quality factor can be associated with the bandwidth:

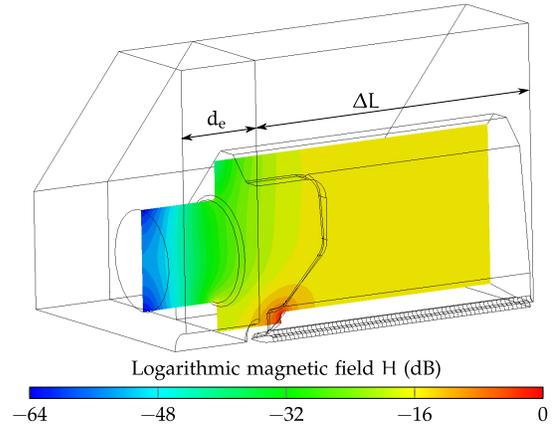
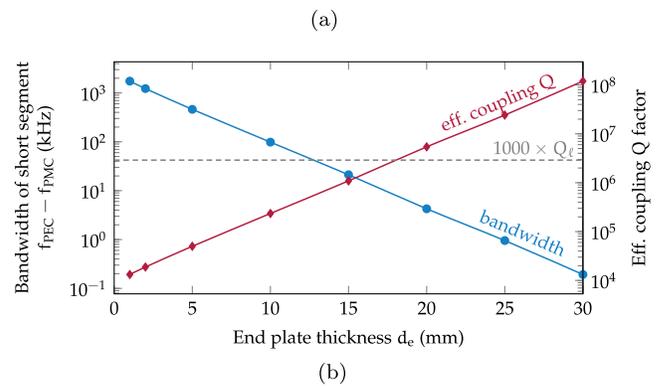

FIG. 20. Rf simulation of a Carbon-RFQ end segment (a) confirming that an end plate thickness of 20 mm is sufficient to decouple the two cavities. Bandwidth $f_{PEC} - f_{PMC}$ and associated effective quality factor are shown in (b).

$$\text{effective } Q = \frac{f_0}{f_{PEC} - f_{PMC}} \cdot \frac{L_{RFQ}}{\Delta L} = 5.3 \times 10^6, \quad (13)$$

where the term $(L_{RFQ}/\Delta L)$ accounts for the fact that only a short segment was simulated by scaling the effective $Q$ to the length of the RFQ cavity. As shown in Fig. 20(b), the effective coupling $Q$ increases exponentially with the end plate thickness $d_e$. For $d_e = 20$ mm and $\Delta L = 7.8$ cm, the bandwidth amounts to $f_{PEC} - f_{PMC} = 4.3$ kHz, and the scaled $Q$ associated with the intercavity coupling reads $5.3 \times 10^6$. This value is more than three orders of magnitude higher than the loaded quality factor $Q_\ell < 3000$ (Sec. IV C), confirming that the decoupling is sufficient and no additional shielding is required.

### C. Beam rematching across the drift

The drift of 4 cm between the output matching plane of RFQ1 and the input matching plane of RFQ2 corresponds to $1.35\beta\lambda$ for the synchronous energy of 2.56 MeV/u at the end of RFQ1. The drift breaks the periodicity of the time-dependent focusing lattice provided by the rf electric quadrupole field. Without any corrections, i.e., ending RFQ1 with a fringe field region (FF) directly after the last accelerating cell, and entering RFQ2 via a radial matcher (RM) [27,45], the beam would be severely mismatched in RFQ2 (Fig. 23). This would result in strong transverse oscillations, phase space filamentation, and particle losses at high energies (Fig. 23, dotted lines).

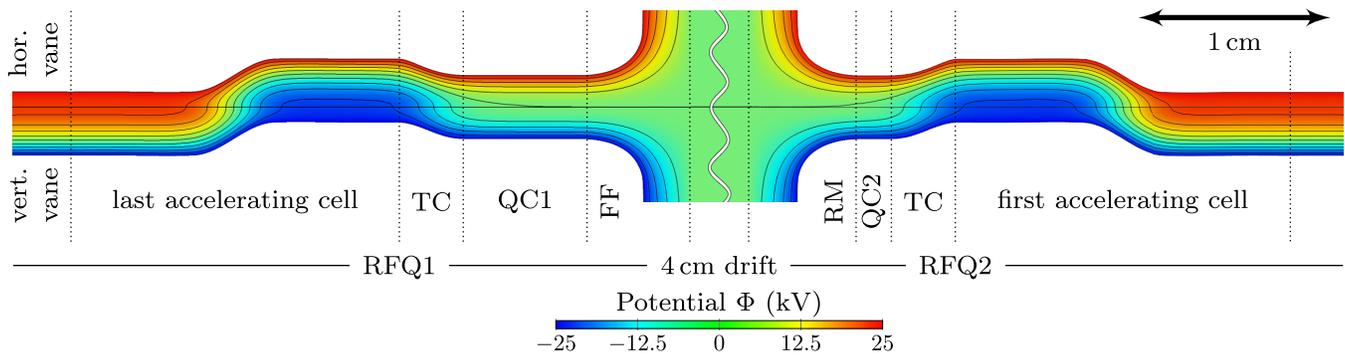

FIG. 21. Field map of the lattice used to transfer the beam from the last accelerating cell of RFQ1 to the first accelerating cell of RFQ2. Both horizontal plane (upper half) and vertical plane (lower half) are shown. After the transition cells (TC), the vanes are extended by the quadrupole cells QC1 and QC2, whose lengths is chosen such that the envelope oscillation in RFQ2 is minimal.

In Ref. [46], Crandall introduced the RFQ transition cell (TC) to smoothly connect an accelerator cell with high modulation ($m > 1$) to a zero-modulation cell ($m = 1$), which from hereon we refer to as quadrupole cell. While the transition cell is commonly used to remove uncertainty in the output energy of an RFQ [27], Crandall pointed out its applicability to divide one long RFQ into two shorter ones. By choosing the length of the quadrupole cell (QC1) after the transition cell, the designer can control the output transverse phase space ellipses at the end of RFQ1. Similarly, RFQ2 starts with an RM much shorter than $\beta\lambda/2$, followed by another quadrupole cell (QC2), followed by a transition cell from $m = 1$ to the previous high modulation, followed by the first accelerating cell of RFQ2 [46]. Figure 21 shows an electrostatic field map computed in COMSOL Multiphysics® [39] of the described lattice.

As RFQ1 and RFQ2 are independent cavities, their rf phases can in principle be chosen arbitrarily. However, the offset is effectively fixed by the longitudinal synchronism requirements: $\Delta\phi = \Delta z/(\beta\lambda)$, where $\Delta z$ is the distance between the end of the last accelerating cell of RFQ1 and the start of the first accelerating cell in RFQ2.

#### 1. Analytic estimate

The two degrees of freedom for the rematching are the lengths of the two quadrupole cells, $L_{QC1}$ and $L_{QC2}$. As an analytic estimate, we considered entering the drift with a symmetric beam, i.e., upright transverse ellipses ($\tilde{\alpha} = 0$). In an ideal RFQ, the beam will be in this configuration at $\phi = 0, \pm 180°$, when the electric field is maximum. The phase at the end of the last accelerating cell of RFQ1 is $\phi_s + 90°$. The required phase advance between this point and the end of the quadrupole cell, where we want $\phi = 180°$, thus reads $90° - \phi_s$. The corresponding length is divided into the transition cell of length $L_{TC}$ and the quadrupole cell of length $L_{QC1}$. Both feature a focusing strength $B(m = 1) = 2.12$ [Eq. (8)] that is significantly stronger than the $B(\text{at cut}) = 1.77$ of the surrounding accelerating cells featuring high modulation (Fig. 12). We take this into account by introducing a simple scaling factor, ignoring the time-dependence of the focusing. The approximate length of the RFQ1 exit quadrupole cell is thus given by

$$L_{QC1} + L_{TC} \approx \beta\lambda \left(\frac{1}{4} - \frac{\phi_s}{360°}\right)\left(\frac{B(\text{at cut})}{B(m=1)}\right). \quad (14)$$

By similar reasoning, we obtain

$$L_{QC2} + L_{TC} \approx \beta\lambda \left(\frac{1}{4} + \frac{\phi_s}{360°}\right)\left(\frac{B(\text{at cut})}{B(m=1)}\right) \quad (15)$$

for the RFQ2 entrance quadrupole cell. These formulas ignore the focusing effect of the short fringe field region and radial matcher at the end of the vanes.

#### 2. Numerical optimization

The final values for $L_{QC1}$ and $L_{QC2}$ were determined by numerical optimization by means of semianalytic field maps generated from the sixteen-term potential function and the tracking code RF-TRACK [41]. The goal was to minimize the maximum transverse rms beam envelopes

$$\sigma_{x,\max} = \max_{z\in\text{RFQ2}}\sigma_x(z), \quad \sigma_{y,\max} = \max_{z\in\text{RFQ2}}\sigma_y(z) \quad (16)$$

in RFQ2. Figure 22 shows a parametric scan of these quantities as functions of the quadrupole cell lengths. Minimizing the expression

$$\sigma_{x,\max} + \sigma_{y,\max} + |\sigma_{x,\max} - \sigma_{y,\max}| \quad (17)$$

delivers the numerical optimum. It differs from the analytic estimate by less than 1 mm $\approx 0.03\beta\lambda$. The plots highlight that the two individual minima for each transverse envelope are offset against each other, corresponding to solutions with asymmetric beams. The combined optimum is a compromise between the two. In expression (17), the third term ensures that the beam is as symmetric as possible.

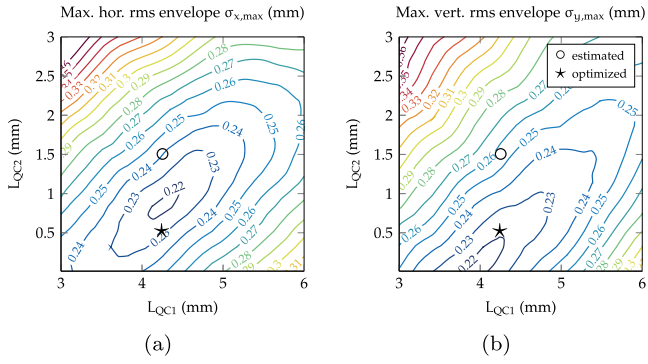

(a)                    (b)

FIG. 22. Scan of the maximum transverse envelopes $\sigma_{x,\max}$ (a) and $\sigma_{y,\max}$ (b) as functions of the lengths of the quadrupole cells $L_{QC1}$ and $L_{QC2}$. The numerically determined optimum parameters ($\star$) shows good agreement with the analytic estimate ($\circ$) obtained from Eqs. (14) and (15).

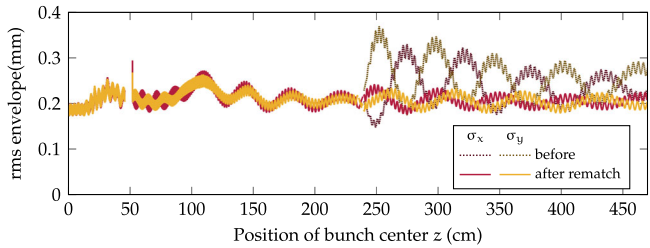

FIG. 23. Beam envelopes of the Carbon-RFQ split into two cavities before and after the rematch.

Figure 23 shows the beam envelopes in both Carbon-RFQ cavities before and after the rematch. The dotted lines show that without any corrections, the beta beating in RFQ2 ($z > 235$ cm) would be unacceptably strong, leading to large beam losses. With the optimum determined above, the transverse envelope beating of the original design (Fig. 15) was almost completely recovered and high-energy particle losses are avoided.

## IV. RF DESIGN

Both methodology and choices of the rf design of the four-vane Carbon-RFQ were adapted from the HF-RFQ [10–14] and PIXE-RFQ [12,15–17]—the obvious exception being the novel dipole mode detuning scheme discussed in Sec. III A. Here, we summarize the remaining rf design considerations, including cavity geometry (Sec. IV A), maximum surface electric field (Sec. IV B), auxiliaries and power couplers (IV C), and thermal simulations (IV D).

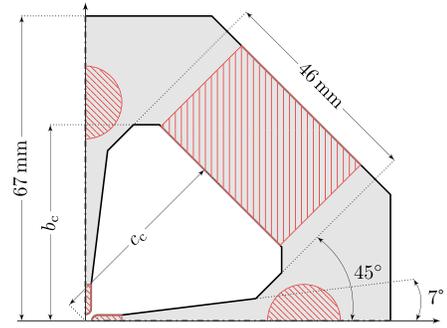

FIG. 24. Quadrant of the Carbon-RFQ cavity cross section with restrictions to the optimization highlighted in red.

### A. Cavity geometry

While the circular quadrant represents the optimum shape for the four-vane cavity in terms of rf losses [27], it is not feasible to construct in practice. Several restrictions apply, which are shown in Fig. 24 (red): The vane tips are of finite thickness and fixed by the beam dynamics design. For mechanical stability and cooling purposes, the vanes itself must be significantly wider than their tips. Between water cooling channels and vacuum domain, at least 4 mm of bulk copper are kept to avoid leakage [44]. The back wall is wide enough such that the auxiliary port apertures intersect solely with a planar surface, hence $b_c = (c_c + 19\,\text{mm})/\sqrt{2}$ for the cavity parameters in Fig. 24.

By simulating individual RFQ cells in Ansys HFSS® [47], an optimum cavity dimension $c_c$ was obtained as a function of the longitudinal coordinate $z$. These simulations furthermore delivered the quadrupole and dipole capacitance $C'$, the dipole cutoff frequency $f_{co}$, and the 2D quality factor $\tilde{Q}_0$, which are shown in Fig. 25. They were used to construct the transmission line (Section III A, Fig. 17), and for the power loss calculation (Sec. IV C).

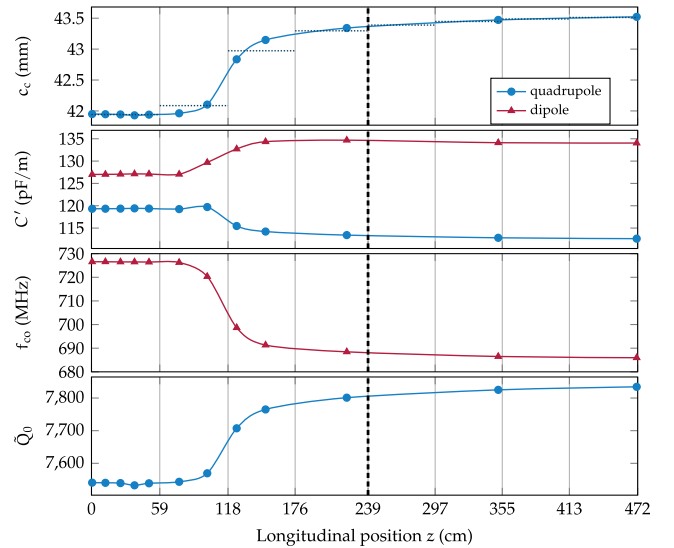

FIG. 25. Rf parameters along the RFQ derived from simulation of individual cells. The vertical lines indicate the approximate dimensions of the $2\times 4$ RFQ modules (intercavity drift is neglected).

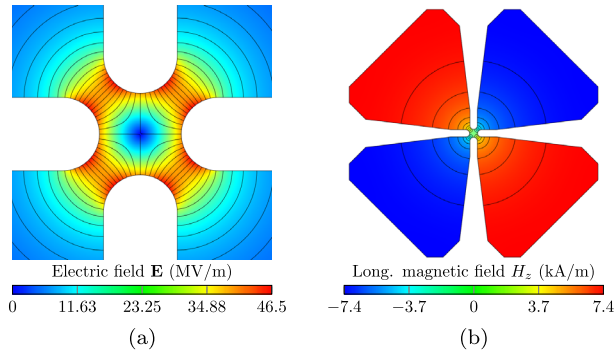

FIG. 26. Electric field lines at the equidistant vane tips of a cell without modulation for the quadrupole mode (a), and magnetic field (b), where the lines connect points of equal amplitude.

Each of the two Carbon-RFQ cavities consists of four individually brazed modules (Fig. 1). The lengths of the copper pieces were chosen such that the intermodule gaps are placed in the center of an RFQ cell, and less than 60 cm to facilitate machining [44]. The optimum $c_c$ with respect to $f_0 = 750$ MHz and maximum $Q_0$ was obtained by averaging the ideal $c_c(z)$-function (Fig. 25, upper plot) over each of the eight modules, such that module-wise constant cross sections were obtained. The cavity cross section and its magnetic field is shown in Fig. 26(b).

### B. Surface electric fields

High surface electric fields may lead to electron emission and sparking, culminating in rf breakdown. Thus, the maximum peak surface electric field $E_{s,\max}$ was studied in detail during the rf design phase.

The 2D cross section simulation, which assumes vane geometry without modulation, delivered an initial $E_{s,\max} = 46.5$ MV/m for a given $V_0 = 50$ kV. This value is only an approximation of the actual maximum field driven by the 3D vane modulation, and in particular small geometric features at the intermodule gaps. Simulation for the PIXE-RFQ have shown that the 3D value can be up to 20% higher than the 2D estimate [16].

Taking the vane modulation into account, a value of 50.5 MV/m was obtained, located in the bunching section of the RFQ [Fig. 27(a)]. One advantage of the trapezoidal vane is the reduced surface field in high-modulation cells in the accelerating section of the RFQ. Although being driven by the sinusoidal step in the center of the cell [Figs. 8(b), 9, 27(b)], the $E_{s,\max}$ is 7% lower compared to a standard cell following the two-term potential function with otherwise same parameters (Fig. 13). However, the overall bottleneck—neglecting any later-introduced gaps—remains in the short high-modulation cells in the bunching section [Fig. 27(a)].

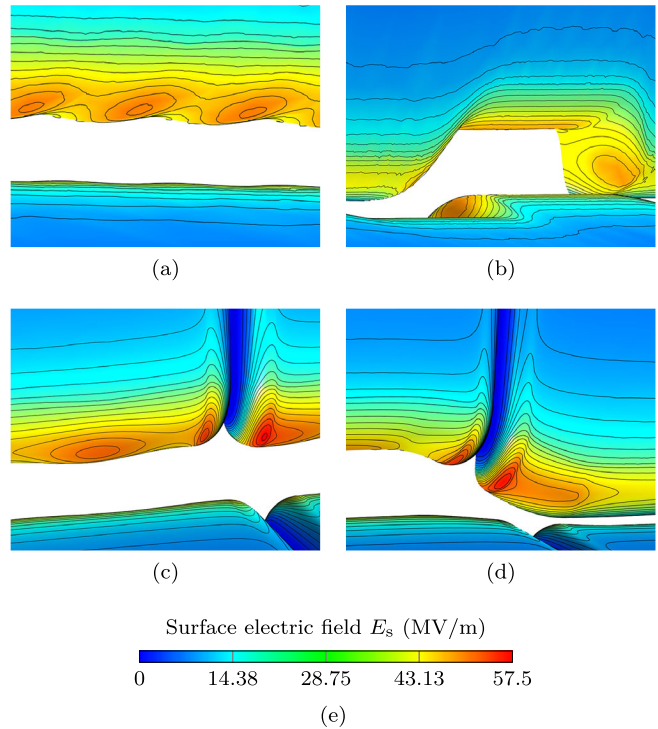

FIG. 27. Surface electric field in normal cells of bunching section (a) and accelerating section (b), as well as in cells interrupted by the inter-module gap in bunching section (c) and accelerating section (d), and (e) legend.

The subdivision of both RFQ cavities into four mechanical, roughly 60 cm long modules introduces small 100 μm gaps into the vane geometry. The gaps are placed in the center of a cell where they have virtually no effect on the beam. In both cavities three gaps are present each, which determine the overall maximum electric field due to the small edge rounding radius. The field at the gap is shown in Fig. 27 for a bunching-section cell (c) and an accelerator-section cell (d). The $E_{s,\max}$ reads 57.5 MV/m, independent of the particular cell. Even though this value exceeds the initially targeted 50.6 MV/m by 14%, we are confident in the rf breakdown resistance of the Carbon-RFQ, as the recently commissioned HF-RFQ features the same inter-module gap geometry at an even higher voltage.

### C. Auxiliaries and power couplers

Each cavity features 32 slug tuners, twelve vacuum pumping ports, and four input power couplers. Additional eight flanges are foreseen for pick-up antenna installation. The auxiliary geometries were optimized for minimum surface losses, and their penetrations into the cavity were adjusted to not perturb the operating mode resonant frequency. Figure 28 shows the surface magnetic field $H_t$ proportional to the surface current density on the tips of tuner (a), vacuum pumping port (b), and power coupler (c).

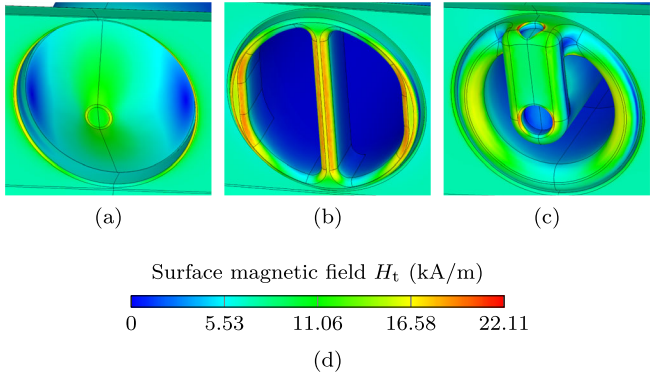

(a)     (b)     (c)

Surface magnetic field $H_\mathrm{t}$ (kA/m)

| 0 | 5.53 | 11.06 | 16.58 | 22.11 |

(d)

FIG. 28. Surface magnetic field $H_\mathrm{t}$ on the tips of tuner (a), vacuum pumping port (b), as well as input power coupler (c), and (d) legend.

TABLE IV. Rf parameters of the Carbon-RFQ cavities.

|  |  | RFQ1 | RFQ2 |  |
|---|---|---|---|---|
| Frequency | $f_0$ | 750 | | MHz |
| Inter-vane voltage | $V_0$ | 50 | | kV |
| Length | $L_\mathrm{RFQ}$ | 235 | 233 | cm |
| Average capacitance | $\langle C' \rangle$ | 117 | 113 | pF/m |
| Stored energy | $W$ | 343 | 329 | mJ |
| Surface power loss | $P_0$ | 244 | 230 | kW |
| Unloaded quality factor | $Q_0$ | 6620 | 6750 | |
| External quality factor, total | $Q_\mathrm{ex}$ | 5000 | | |
| Loaded quality factor | $Q_\ell$ | 2850 | 2870 | |
| Coupling strength | $\beta_\mathrm{c}$ | 1.32 | 1.35 | |
| Number of couplers | | 4 | | |
| External quality factor, per coupler | | 20000 | | |
| Power per coupler ($\beta_\mathrm{c} = 1$) | | 61 | 60 | kW |

### 1. Tuning system

The tuning system consists of 32 copper slug tuners for each of the approximately 240 cm $= 6\lambda$ long RFQ cavities, corresponding to 1.3 tuners/$\lambda$/quadrant. Although $6\lambda$ is considered close to the limit of field stabilization using only piston tuners, we are confident in the capabilities of this tuning system considering the available spectral margins (Fig. 19). The slightly shorter ($5\lambda$) HF-RFQ has been successfully tuned with this configuration [13]. Each slug features a movement range of $\pm 11$ mm with respect to nominal, such that the available frequency tuning range amounts to $\pm 11$ MHz. Because of the more inhomogenous capacitance profile (Fig. 25), we expect the tuning of RFQ1 to be more challenging, as the field is more tilted when the tuners are in nominal position. However, the necessary corrections are well within the piston movement range.

### 2. RFQ power and couplers

The power loss of both RFQ cavities was calculated by means of the method presented in Ref. [16], which has been shown to agree with a full 3D eigenmode computation up to an error of $\approx 1\%$. Each RFQ was considered as a concatenation of segments that coincide with the ports of tuners, pumping ports, couplers, and ends.[2] The $n$th segment contributes an rf power loss of $P_n = \omega_0 W_n/Q_{0,n}$, where the stored energy $W_n$ is obtained by integrating the capacitance profile (Fig. 25, second plot). $Q_{0,n}$ was calculated both from the contribution of the vane modulation (Fig. 25, bottom plot), and from 3D eigenmode computation of individual segments. The obtained rf quantities are summarized in Table IV.

Each cavity will be supplied by four coaxial magnetic loop couplers. While each coupler is strongly undercoupled individually, combined they are overcoupled by a margin of 30% ($\beta_\mathrm{c} = Q_0/Q_\mathrm{ex} = 1.32$ and 1.35, respectively). The couplers will be mounted on rotatable flanges, allowing to fine-tune the coupling with the help of rf measurements after assembly. In the case of critical coupling ($\beta_\mathrm{c} = 1$), each coupler supplies a (peak) power of 60 kW. High-power tests with the HF-RFQ have demonstrated that the coupler can transmit up to 100 kW. More details on coupler geometry and vacuum window are found in Ref. [16].

---
[2]The pick-up antennas are neglected in this estimation because of their marginal contribution.

### D. Thermal simulation

During operation, rf losses act as a heat source on the copper structure. The resulting temperature increase causes deformation of the cavity and detuning of the resonant frequency. Thermal and structural simulations were conducted using COMSOL Multiphysics® [39] to study this behavior and determine cooling system requirements.

Because of the low duty cycle of $d = T_\mathrm{pls} f_\mathrm{rep} = 0.1\%$, the cooling system must dissipate only 240 W per RFQ. It rather serves for frequency stabilization. In Fig. 29, the frequency shift $\Delta f$ is shown as a function of the duty cycle $d$ and the cooling water temperature $T_\mathrm{w}$. The average water speed in the cooling channels amounts to 1 m/s, corresponding to a convection coefficient of 3900 W/m$^2$/K. Two configurations are compared: a design with two channels per vane, used by the HF-RFQ and the PIXE-RFQ [15,16], and a new design with only one channel per vane. The plot highlights that the difference between the two options is only marginal, such that the mechanically simpler design with one channel per vane was selected.

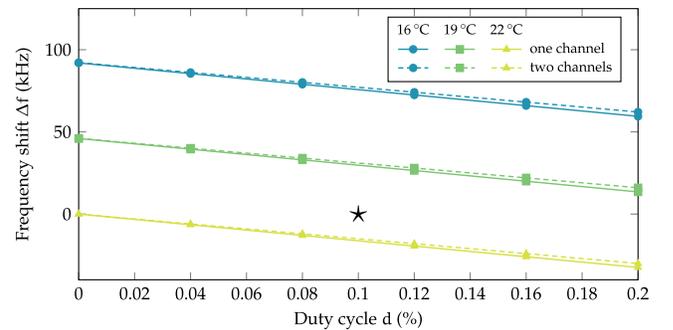

FIG. 29. Resonance frequency shift $\Delta f$ of the RFQ cavity subject to heat-induced deformation as a function of duty cycle $d$ and cooling water temperature $T_\mathrm{w}$ for the cases of one and two cooling channels per vane. At nominal operation, the heating is compensated by reducing $T_\mathrm{w}$ by 1 K ($\star$).

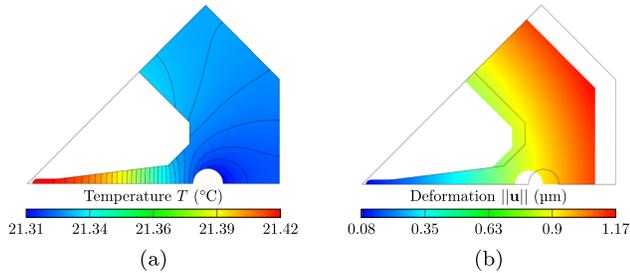

FIG. 30. Temperature $T$ (a) and deformation $\mathbf{u}$ (b) in the 2D cavity subject to rf power losses. The distributions are shown for the nominal operating point with 0.1% duty cycle and 21 °C cooling water temperature (★ in Fig. 29).

Figure 29 visualizes the available frequency tuning range for a given water temperature range: $\partial f/T_w = 15.3$ kHz/K. Furthermore, the frequency is a linearly decreasing function of the duty cycle with $\partial f/\partial d = -151$ kHz/%. Thus, to recover the design resonance frequency at nominal operation, the cooling water temperature must be reduced by approximately 1 K (★ in Fig. 29). Figure 30 shows the distributions of copper temperature $T$ (a) and the deformation field $\mathbf{u}$ (b) in the 2D cross section for this point of operation. Because of the temperature gradient in the vane, the vane tips move closer toward each other, increasing the capacitance $C$ and thus lowering the frequency with increasing duty cycle. This is compensated for by decreasing cavity temperature and thus volume, hereby lowering the inductance $L$, such that with $\partial f/f = -(\partial L/L + \partial C/C)/2$ the design resonant frequency is restored.

## V. CONCLUSION AND OUTLOOK

A 750 MHz RFQ with trapezoidal vanes for fully stripped carbon ions was designed at CERN, to be integrated as part of a carbon ion therapy facility. It represents the first rf accelerating structure of the recently proposed "bent linac." All of its components were optimized to match the requirements of compactness, low power consumption, and simplification of construction in view of industrialization.

Considering both high-transmission and compact options, as well as different output energies, a compact RFQ with standard two-term potential vanes was designed, which accelerates particles with a charge-to-mass ratio of 1/2 to 5 MeV/u. In this configuration, particle losses at energies higher than a few 10 keV were minimized.

The standard-vane RFQ was then redesigned to feature trapezoidal vanes, reducing power consumption and length by 15%. A novel semianalytic technique using sixteen terms of the multipole potential function was used to determine the vane shape in a cell-by-cell manner. The beam dynamics design was validated by particle tracking simulation with both semianalytic and FEM-based field maps. These simulation showed that the augmented potential function can describe both longitudinal and transverse beam dynamics in close agreement with an FEM simulation of the real vane geometry. As only one specific trapezoidal electrode type was considered and some discrepancy between potential function and FEM field map was observed in the longitudinal plane (although still well within requirements), future studies should challenge the applicability of the sixteen-term potential function in other examples, particularly with respect to other electrode types. Should the sixteen multipoles prove insufficient, more coefficients could be added. In the longer term, a database providing multipole data for RFQ electrodes of various types could be created, as has been done for the PARMTEQ codes for conventional RFQ electrodes. It would provide a highly useful tool for rapidly designing RFQs with trapezoidal vanes.

The rf design of the Carbon-RFQ has been completed, focusing on minimized power loss and simplified construction. The geometries of cavity, auxiliaries, and power couplers were determined. A major feature is the subdivision into two fully decoupled rf cavities. The rematching of the beam across the drift was accomplished by two quadrupole vane extensions, which were optimized for minimum beam envelope oscillations. The length of each of the two RFQ cavities was used as a manner of dipole mode detuning. By means of this novel method, maximum spectral margins were achieved without the use of any other dedicated detuning technique. The rf design was validated by accurately simulating the maximum electric field on the most critical surfaces, as well as thermal simulations.

In the framework of a collaboration between CERN and CIEMAT (Centro de Investigaciones Energéticas, Medioambientales y Tecnológicas), the Carbon-RFQ will be built as part of a demonstrator of the low-energy part of linacs for carbon ion therapy. The demonstrator will be composed of a copy of the TwinEBIS source, a LEBT, the Carbon-RFQ, and some further structure (studies presently ongoing for the choice of the accelerating structure following the RFQ). It has the purpose of validating the beam quality at the end of the low energy section, which is often the most critical part of hadron linacs, defining the beam characteristics that are then propagated along the whole accelerator.

## ACKNOWLEDGMENTS

This work has been supported by the CERN Knowledge Transfer Group and the Wolfgang Gentner Programme of the German Federal Ministry for Education and Research (BMBF, grant no. 05E15CHA). We would like to thank

[1] U. Amaldi, S. Braccini, and P. Puggioni, High frequency linacs for hadrontherapy, Rev. Accel. Sci. Technol. **02**, 111 (2009).

[2] A. Degiovanni, P. Stabile, and D. Ungaro, LIGHT: a linear accelerator for proton therapy, in *Proceedings of NAPAC2016, Chicago, USA* (JACoW, Geneva, Switzerland, 2016).

[3] C. Ronsivalle, M. Carpanese, C. Marino, G. Messina, L. Picardi, S. Sandri, E. Basile, B. Caccia, D. Castelluccio, E. Cisbani et al., The TOP-IMPLART project, Eur. Phys. J. Plus **126**, 68 (2011).

[4] P. Ostroumov, L. Faillace, A. Goel, S. Kutsaev, and B. Mustapha, Compact Carbon Ion Linac, in *2nd North American Particle Accelerator Conference (NAPAC2016)* (JACoW, Geneva, Switzerland, 2016), pp. 61–63.

[5] S. Verdú-Andrés, U. Amaldi, and Á. Faus-Golfe, CABOTO, a high-gradient linac for hadrontherapy, Journal of Radiation Research **54**, i155 (2013).

[6] S. Benedetti, High-gradient and high-efficiency linear accelerators for hadron therapy, Ph.D. thesis, Ecole Polytechnique Fédérale de Lausanne, 2018.

[7] V. Bencini, Design of a novel linear accelerator for carbon ion therapy, Ph.D. thesis, Sapienza Università di Roma (2020), available as Report No. CERN-THESIS-2019-301.

[8] M. Breitenfeldt, R. Mertzig, J. Pitters, A. Shornikov, and F. Wenander, The TwinEBIS setup: machine description, Nucl. Instrum. Methods Phys. Res., Sect. A **856**, 139 (2017).

[9] H. Pahl, V. Bencini, M. Breitenfeldt, A. G. Costa, A. Pikin, J. Pitters, and F. Wenander, A low energy ion beamline for TwinEBIS, J. Instrum. **13**, P08012 (2018).

[10] M. Vretenar, A. Dallocchio, V. A. Dimov, M. Garlaschè, A. Grudiev, A. M. Lombardi, S. Mathot, E. Montesinos, and M. Timmins, A compact high-frequency RFQ for medical applications, in *27th Linear Accelerator Conference (LINAC2014)* (JACoW, Geneva, Switzerland, 2014), pp. 935–938.

[11] A. M. Lombardi, E. Montesinos, M. Timmins, M. Garlaschè, A. Grudiev, S. Mathot, V. Dimov, S. Myers, and M. Vretenar, Beam dynamics in a high-frequency RFQ, in *6th International Particle Accelerator Conference (IPAC2015)* (JACoW, Geneva, Switzerland, 2015), pp. 2408–2412.

[12] M. Vretenar, E. Montesinos, M. Timmins, M. Garlaschè, A. Grudiev, S. Mathot, B. Koubek, V. Dimov, A. M. Lombardi, and D. Mazur, High-frequency compact RFQs for medical and industrial applications, in *28th Linear Accelerator Conference (LINAC2016)* (JACoW, Geneva, Switzerland, 2016), pp. 704–709.

[13] B. Koubek, A. Grudiev, and M. Timmins, rf measurements and tuning of the 750 MHz radio frequency quadrupole, Phys. Rev. Accel. Beams **20**, 080102 (2017).

[14] V. A. Dimov, M. Caldara, A. Degiovanni, L. S. Esposito, D. A. Fink, M. Giunta, A. Jeff, A. Valloni, A. M. Lombardi, S. J. Mathot, and M. Vretenar, Beam commissioning of the 750 MHz proton RFQ for the LIGHT prototype, in *9th International Particle Accelerator Conference (IPAC2018)* (JACoW, Geneva, Switzerland, 2018), pp. 658–660.

[15] H. W. Pommerenke, A. Bilton, A. Grudiev, A. M. Lombardi, S. Mathot, E. Montesinos, M. Timmins, M. Vretenar, and U. van Rienen, RF design of a high-frequency RFQ linac for PIXE analysis, in *29th Linear Accelerator Conference (LINAC2018)* (JACoW, Geneva, Switzerland, 2018), pp. 822–825.

[16] H. W. Pommerenke, V. Bencini, A. Grudiev, A. M. Lombardi, S. Mathot, E. Montesinos, M. Timmins, U. van Rienen, and M. Vretenar, rf design studies on the 750 MHz radio frequency quadrupole linac for proton-induced x-ray emission analysis, Phys. Rev. Accel. Beams **22**, 052003 (2019).

[17] S. Mathot et al., The CERN PIXE-RFQ, a transportable proton accelerator for the machina project, Nucl. Instrum. Methods Phys. Res., Sect. B **459**, 153 (2019).

[18] T. P. Wangler, Lumped-circuit model of four-vane RFQ resonator, Los Alamos National Lab., NM (USA) Tech. Report, 1984.

[19] W. D. Kilpatrick, Criterion for vacuum sparking designed to include both RF and DC, Rev. Sci. Instrum. **28**, 824 (1957).

[20] T. Boyd, Jr., Kilpatrick's criterion, Los Alamos National Lab., NM (USA), Technical Report AT-1:82-28, 1982.

[21] K. R. Crandall and T. P. Wangler, PARMTEQ—a beam dynamics code for the RFQ linear accelerator, in *AIP Conference Proceedings* (AIP, New York, 1988), pp. 22–28.

[22] K. R. Crandall, R. H. Stokes, and T. P. Wangler, RF quadrupole beam dynamics design studies, in *10th Linear Accelerator Conference (LINAC1979)* (JACoW, Geneva, Switzerland, 1979), pp. 205–216, https://inspirehep.net/literature/144542.

[23] S. Benedetti, A. Latina, and A. Grudiev, Design of a 750 MHz IH structure for medical applications, in *Proceedings of LINAC2016* (JACoW, Geneva, Switzerland, 2017), pp. 240–243.

[24] L. Picardi, P. Nenzi, A. Ampollini, C. Ronsivalle, F. Ambrosini, M. Vadrucci, G. Bazzano, and V. Surrenti, Experimental results on SCDTL structures for protons, in *Proceedings of IPAC2014, Dresden, Germany* (JACoW Publishing, Geneva, Switzerland, 2014), pp. 3247–3249, https://doi.org/10.18429/JACoW-IPAC2014-THPME016.

[25] A. Degiovanni, M. Esposito, Y. Fusco, F. Salveter, H. Pavetits, A. Marraffa, C. Mellace, P. Stabile, D. A.

Murciano, K. Stachyra et al., Status of the commissioning of the LIGHT prototype, in *9th International Particle Accelerator Conference (IPAC2018)* (JACoW, Geneva, Switzerland, 2018), pp. 425–428.

[26] A. S. Plastun and P. N. Ostroumov, Practical design approach for trapezoidal modulation of a radio frequency quadrupole, Phys. Rev. Accel. Beams 21, 030102 (2018).

[27] T. P. Wangler, *RF linear accelerators*, 2nd ed. (John Wiley & Sons, New York, 2008).

[28] K. R. Crandall, Effects of vane tip geometry on the electric fields in radio frequency quadrupole linacs, Los Alamos National Laboratory Tech. Report, 1983.

[29] K. R. Crandall, Computation of charge distribution on or near equipotential surfaces, Los Alamos Scientific Laboratory Tech. Report, 1996.

[30] R. Ferdinand, R. Duperrier, J.-M. Lagniel, P. Mattei, and S. Nath, Field description in an RFQ and its effect on beam dynamics, in *19th Linear Accelerator Conference (LINAC1998)* (1998), https://cds.cern.ch/record/740859.

[31] A. Letchford and A. Schempp, A comparison of 4-rod and 4-vane RFQ fields, in *6th European Particle Accelerator Conference (EPAC1998)* (JACoW, Geneva, Switzerland, 1998), pp. 1204–1206, https://accelconf.web.cern.ch/e98/PAPERS/THP11E.PDF.

[32] O. K. Belyaev, O. V. Ershov, I. G. Maltsev, V. B. Stepanov, S. A. Strekalovskikh, V. A. Teplyakov, and A. V. Zherebtsov, IHEP experience on creation and operation of RFQs, in *20th Linear Accelerator Conference (LINAC2000)* (SLAC, Menlo Park, CA, USA, 2000), pp. 259–261.

[33] B. Mustapha, A. A. Kolomiets, and P. N. Ostroumov, Full 3D modeling of a radio-frequency quadrupole, in *25th Linear Accelerator Conference (LINAC2010)* (JACoW, Geneva, Switzerland, 2010), pp. 542–544, https://inspirehep.net/literature/1363849.

[34] A. S. Plastun and A. A. Kolomiets, RFQ with improved energy gain, in *26th Linear Accelerator Conference (LINAC2012)* (JACoW, Geneva, Switzerland, 2012), pp. 966–968.

[35] B. Mustapha, A. A. Kolomiets, and P. N. Ostroumov, Full three-dimensional approach to the design and simulation of a radio-frequency quadrupole, Phys. Rev. Accel. Beams 16, 120101 (2013).

[36] C. Li, Y. He, and Z. Wang, Optimization design of the RFQ trapezoidal electrode, in *13th International Conference on Heavy Ion Accelerator Technology (HIAT2015)* (2015).

[37] Y. Iwashita, Y. Fuwa, and R. A. Jameson, RFQ vane shapes for efficient acceleration, in *28th Linear Accelerator Conference (LINAC2016)* (JACoW, Geneva, Switzerland, 2016), pp. 581–583.

[38] H. P. Li, Z. Wang, Y. R. Lu, Q. Y. Tan, M. J. Easton, K. Zhu, S. Liu, and M. Y. Han, Novel deuteron RFQ design with trapezoidal electrodes and double dipole four-vane structure, J. Instrum. 15, T02006 (2020).

[39] COMSOL AB, COMSOL Multiphysics®, version 5.4 (2018).

[40] A. Perrin, J. Amand, and T. Mütze, Travel v4. 06, user manual, CERN internal note, 2003.

[41] A. Latina, RF-Track: beam tracking in field maps including space-charge effects, features and benchmarks, in *28th Linear Accelerator Conference (LINAC2016)* (JACoW, Geneva, Switzerland, 2016), pp. 104–107.

[42] F. Grespan, A. Pisent, and A. Palmieri, Dipole stabilizers for a four-vane high current RFQ: theoretical analysis and experimental results on a real-scale model, Nucl. Instrum. Methods Phys. Res., Sect. A 582, 303 (2007).

[43] Computer Simulation Technology, CST Studio Suite®, release 2018 (2018).

[44] S. Mathot (private communication).

[45] K. R. Crandall, RFQ radial matching sections and fringe fields, in *1984 Linear Accelerator Conference (LINAC84)* (JACoW, Geneva, Switzerland, 1984), pp. 109–111, https://accelconf.web.cern.ch/l84/papers/tup0009.pdf.

[46] K. R. Crandall, Ending RFQ vanetips with quadrupole symmetry, in *17th Linear Accelerator Conference (LINAC1994)* (JACoW, Geneva, Switzerland, 1994), pp. 227–229, https://accelconf.web.cern.ch/l94/papers/mo-69.pdf.

[47] ANSYS Inc., Electronics Desktop®, version 2019 r1 (2019).